\documentclass[twocolumn]{aastex63}

\newcommand{\x}{X-ray}

\newcommand{\chandra}{\textit{Chandra}}

\newcommand{\xmm}{\textit{XMM-Newton}}

\newcommand{\nustar}{\textit{NuSTAR}}

\newcommand{\exi}{\begin{equation}}
\newcommand{\exo}{\end{equation}}

\usepackage[utf8]{inputenc}
\usepackage{multirow}
\usepackage{tablefootnote}
\usepackage{wrapfig}
\usepackage{xurl}
\shorttitle{Hydrostatic Mass of A478}
\shortauthors{Potter}

\begin{document}

\title{The Hydrostatic Mass of A478: Discrepant Results From \chandra,\ \nustar,\ and \xmm}

\author[0009-0003-7738-9173]{Cicely Potter}
\affiliation{Department of Physics \& Astronomy, The University of Utah, 115 South 1400 East, Salt Lake City, UT 84112, USA}

\author[0000-0002-3132-8776]{Ay\c{s}eg\"{u}l T\"{u}mer}
\affiliation{Kavli Institute for Astrophysics and Space Research, Massachusetts Institute of Technology, 77 Massachusetts Avenue, Cambridge, MA 02139, USA}
\affiliation{Department of Physics \& Astronomy, The University of Utah, 115 South 1400 East, Salt Lake City, UT 84112, USA}

\author[0000-0001-8468-9164]{Qian H. S. Wang}
\affiliation{Department of Astronomy, University of Maryland, College Park, MD 20742, USA}

\author[0000-0001-9110-2245]{Daniel R. Wik}
\affiliation{Department of Physics \& Astronomy, The University of Utah, 115 South 1400 East, Salt Lake City, UT 84112, USA}

\author[0000-0003-0791-9098]{Ben J. Maughan}
\affiliation{H. H. Wills Physics Laboratory, University of Bristol, Tyndall Ave, Bristol BS8 1TL, UK}

\author[0000-0002-4962-0740]{Gerrit Schellenberger}
\affiliation{Center for Astrophysics, Harvard \& Smithsonian, 60 Garden Street, Cambridge, MA 02138, USA}

\begin{abstract}
Galaxy clusters are the most recently formed and most massive, gravitationally bound structures in the universe. The number of galaxy clusters formed is highly dependent on cosmological parameters, such as the dark matter density, $\sigma_8$, and $\Omega_m$. The number density is a function of the cluster mass, which can be estimated from the density and temperature profiles of the intracluster medium (ICM) under the assumption of hydrostatic equilibrium. The temperature of the plasma, hence its mass, is calculated from the X-ray spectra. However, effective area calibration uncertainties in the soft band result in significantly different temperature measurements from various space-based X-ray telescopes. \nustar\ is potentially less susceptible to these issues than \chandra\ and \xmm,\ having larger effective area, particularly at higher energies, enabling high precision temperature measurements. In this work, we present analyses of \chandra,\ \nustar,\ and \xmm\ data of Abell 478 to investigate the nature of this calibration discrepancy. We find that \nustar\ temperatures are on average $\sim$11\% lower than that of \chandra,\ and \xmm\ temperatures are on average $\sim$5\% lower than that of \nustar.\ This results in a \nustar\ mass at $r_{2500,Chandra}$ of $M_{2500,NuSTAR}=3.39^{+0.07}_{-0.07}\times10^{14}$ $M_{\odot}$, which is $\sim$10\% lower than that of $M_{2500,Chandra}$ and $\sim$4\% higher than $M_{2500,XMM-Newton}$.

\end{abstract}


\section{introduction}

Galaxy clusters form hierarchically from smaller, virialized structures at the peaks of the initial matter density fluctuations of the universe. 
Thus galaxy cluster number density is strongly dependent on the matter density of the universe, $\Omega_{m}$, and the amplitude of the power spectrum, $\sigma_{8}$, as well as the evolution of dark energy.
The number density is a function of the cluster mass, meaning cluster mass can be used to constrain cosmological parameters (for example, \citet{vikhlinin2010}, \citet{Burenin2012}, \citet{bartalucci2018}, \citet{Ettori2019}, \citet{ferragamo2021}).

Cluster mass can be calculated from the density and temperature profiles of the intracluster medium (ICM) under the assumption of hydrostatic equilibrium. These hydrostatic cluster mass measurements require a bias factor to be consistent with the best-fit Planck base-$\Lambda$CDM cosmology as measured by cosmic microwave background anisotropies. While the bias factor of $(1-b) = 0.62\pm0.03$ found by \citet{planck2021} is within 1$\sigma$ of weak lensing measurements, it is at the lower end.

X-ray spectra are ideal for measuring temperature; these temperature measurements, however, critically depend on the broadband calibration of a telescope's effective area, or collecting area as a function of energy. The effective area will be less than the geometric area of the mirrors %detectors?
due to factors such as the quantum efficiency, vignetting, molecular contamination, and optical blocking filter transmission. Different X-ray telescopes report different temperatures for the same clusters, resulting in the masses differing by $\sim$10$\%$. This is suspected to be a calibration issue.

For a large sample of bright, nearby galaxy clusters \citet{schellenberger2014} find that, at a cluster temperature of 10 keV, temperatures measured by the \xmm\ European Photon Imaging Camera silicon pn-junction CCD (EPIC-PN) are on average $23\%$ lower than temperatures measured by \chandra\ Advanced CCD Imaging Spectrometer (ACIS). This is due to the effective area calibration uncertainties, mainly in the 0.7--2 keV band, between \xmm\ and \chandra\ that have been revealed through the model-independent stacked residuals ratios (see also \citet{Nevalainen10} and \citet{Kettula13}). The difference between \chandra\ and \xmm\ increases with cluster temperature, thus hotter clusters are ideal for investigating this difference further, and \nustar,\ with its harder response, is able to measure hotter temperatures more precisely.

\xmm\ EPIC detectors MOS1, MOS2, and PN show disagreement as well, where systematically lower temperatures are measured with PN. When comparing the hard and soft band temperatures of one instrument, \citet{schellenberger2014} found that Chandra ACIS temperatures are consistent with each other, while XMM-Newton EPIC are not. However, depending on the level of multiphase gas along the line of sight, a perfectly calibrated instrument is expected to show some level of deviation between the soft and hard band temperatures. Recent corrections to the effective areas of all three of \xmm’s\ EPIC cameras were made with the intent to bring them into better agreement with \nustar\footnote{\url{https://xmmweb.esac.esa.int/docs/documents/CAL-SRN-0388-1-4.pdf}}. These corrections only apply to the 3.0--12.0 keV band and are found to be successful in improving agreement between EPIC-PN and \nustar's\ Focal Plane Modules A and B (FPMA and FPMB). 
%This revision could potentially resolve the broadband discrepancy?

\nustar\ is potentially less susceptible to calibration issues than \xmm\ and \chandra,\ having greater sensitivity at higher energies, where the effective area is relatively constant and the exponential turnover of the bremsstrahlung continuum is more prominent given its bandpass for hotter clusters. The exponential turnover provides the most constraining power on temperature in low resolution spectra. 
% this is said above, doesn't really need to be stated again here
%\todo{and the disagreement between \chandra\ and \xmm\ increases with cluster temperature}. 
In a recent \nustar\ effective area calibration update, stray light observations of the Crab Nebula were used, allowing a more accurate measurement of the vignetting function; stray light observations bypass the optics, thus avoiding any degeneracy with the multilayer insulation and Be window. In addition to effective area changes, these updates bring FPMA and FPMB flux into better agreement with each other, increase the measured flux by $5$--$15\%$ (depending on off-axis angle), and make more accurate high energy and off-axis angle corrections \citet{madsen2021}. \citet{wallbank22} find that, for a sample of 8 galaxy clusters, \nustar\ temperatures are $\sim$10\% and $\sim$15\% lower than \chandra\ temperatures in the broad and hard bands respectively.

Abell 478 (A478) is a nearby (z~=~0.0856), massive, relaxed, cool-core galaxy cluster with $r_{2500,Chandra}$ \footnote{$r_{\Delta}$ corresponds to the radius of a point on a sphere whose density equals $\Delta$ times the critical density of the background universe, $\rho_{c}$(z), at the cluster redshift where $\Delta$ is the density contrast.} $=589 \pm 14$ kpc and $r_{500,Chandra} \sim1514$ kpc.

\citet{vik06_2} measure the spectroscopic temperature of A478 with \chandra\ data to be $7.95 \pm 0.14$~keV, and \citet{arnaud2005} measure it with \xmm\ data to be $7.05 \pm 0.12$~keV. This results in masses of 4.12 $\pm$ 0.26 $\times$ $10^{14}$ $M_{\odot}$ and 3.12 $\pm$ 0.31 $\times$ $10^{14}$ $M_{\odot}$ at $r_{2500}$ for \chandra\ and \xmm,\ respectively. This is a difference of 11\% between spectroscopic temperatures and 26\% between masses. This disagreement is partially due to differences in analysis, such as the regions the spectra are extracted from or the choice of spectral models and their parameters. For example, some of the difference may have since been resolved with the \chandra\ CALDB update from version 3 to 4. While some differences are unavoidable due to inherent instrument properties, we attempt to provide analyses of the data from \chandra,\ \nustar,\ and \xmm\ that differ in method as little as possible.

%Arnaud: mekal and wabs

%Vikhlinin: absorbed(?) mekal, fix abundance outside 500 kpc to 0.33,

Throughout this paper, we assume $\Lambda$CDM cosmology with {\it H$_{0}$} = 71 km s$^{-1}$ Mpc$^{-1}$, $\Omega_{M}$ = 0.3, $\Omega_{\Lambda}$ = 0.7. According to these assumptions, at the cluster redshift, a projected intracluster distance of 100 kpc corresponds to an angular separation of $\sim$61$\arcsec$. All uncertainties are quoted at the 68\% confidence levels unless otherwise stated.

The paper is organized as follows: description of the data, data reduction processes, and background assessment for \chandra,\ \nustar,\ and \xmm\ are presented in Section~\ref{sec:reduction}. The methods used for the analyses of the cluster data and the results are presented in Section~\ref{sec:analysis}. In Section~\ref{sec:discussion}, we discuss our findings.

%\AT{I would combine data analysis and results sections, maintaining a good flow.}

\section{Data Reduction}\label{sec:reduction}

%\AT{Here, describe the data first. Observation IDs, how long was the exposure time, and when it was taken, which detectors were used. Then use two subsections: one for NuSTAR and one for Chandra reductions(I'd start with NuSTAR). The subsections you currently have, i.e.; Extracting Temperature and Density Profiles and Crosstalk Correction belongs to the data analysis. Reduction is only for describing how the data is cleaned and filtered.}

\begin{table*}
\centering
\begin{tabular}{ ccccccc } 
\multicolumn{7}{c}{\large{\textbf{Observation Log}}}\\
\hline
\hline
 & & Date & RA & Dec & Exposure & \\
 & Observation ID & (yyyy-mm-dd) & (J2000) & (J2000) & (ks) & PI\\
\hline
\chandra\ & 1669 (ACIS-S) & 2001-01-27 & 63.36 & 10.44 & 42.9 & Murray\\ 
 & 6102 (ACIS-I) & 2004-09-13 & 63.37 & 10.49 & 10.5 & Allen\\
\hline
\nustar\ & 70660002002 & 2020-09-29 & 63.35 & 10.45 & 207.9 (FPM(A+B)) & Wik\\ 
 & 70660002004 & 2020-09-29 & 63.35 & 10.45 & 260.7 (FPM(A+B)) & Wik\\
\hline
\xmm\ & 109880101 & 2002-02-15 & 63.38 & 10.47 & 57.7 (MOS1)/94.2 (MOS2)/62.6 (PN) & Brinkman\\
\hline
\end{tabular}
\caption{Observation log of the \chandra,\ \nustar,\ and \xmm\ data used for analysis. The exposure shown is that of the observations before filtering and data reduction.}
\label{data}
\end{table*}

\begin{figure*}[ht]
 \centering
 \includegraphics[scale=0.35]{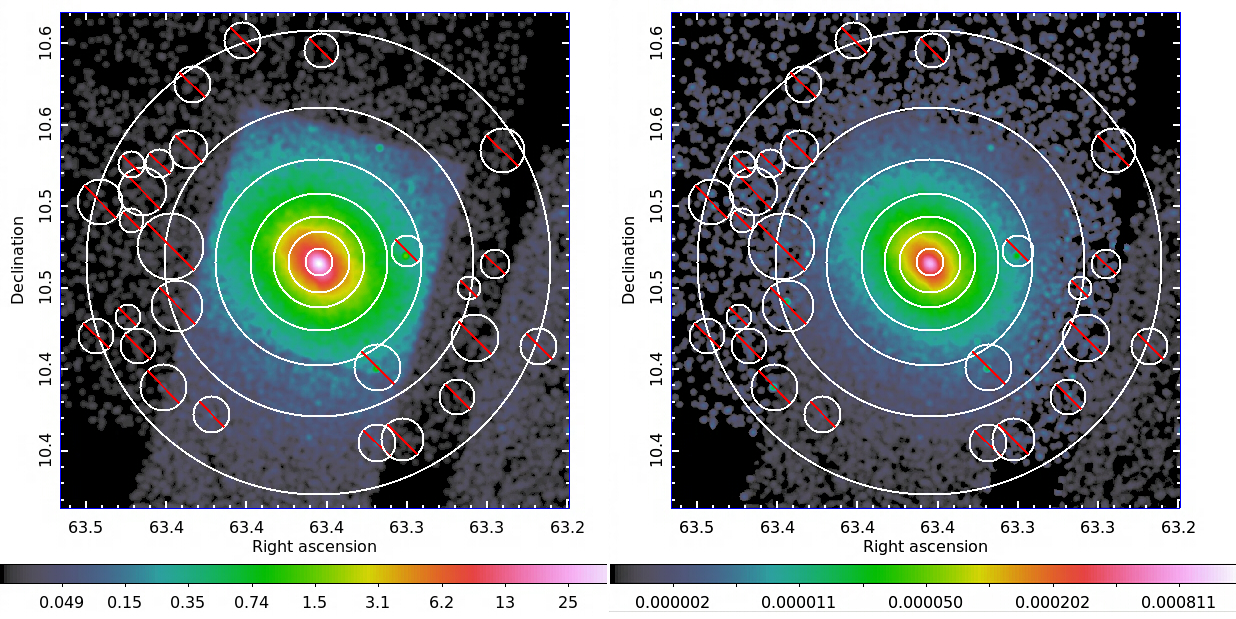}
 \caption{The raw (Left) and background subtracted, exposure corrected (Right) 0.8--2 keV counts images of the \chandra\ observations. The spectrum was extracted from each of the annuli shown; these spectra were fit to get the observed temperature profile. The excluded regions are the point sources manually selected from a 0.5--4.0 keV image. The radii of the annuli are 30\arcsec, 68\arcsec, 101\arcsec, 152\arcsec, 228\arcsec, 342\arcsec, and 513\arcsec.}
 \label{chandra_images}
\end{figure*}

\subsection{Chandra}
In this work we used two sets of \chandra\ data consisting of one ACIS-S and one ACIS-I pointing (see Table \ref{data}). For the \chandra\ data reduction, we use HEASoft\footnote{\url{https://heasarc.gsfc.nasa.gov/lheasoft/}} version 6.30.1, CIAO\footnote{\url{https://cxc.cfa.harvard.edu/ciao/}} version 4.12, and CALDB version 4.7.6. The python code {\tt acisgenie}\footnote{\url{https://gitlab.com/qwq/acisgenie}} was used to generate scripts to run the standard CIAO data filtering commands.

First, any linear streaks caused by flaws in readout were removed with {\tt destreak}. Bad pixels both from the framestore and the observation itself, including afterglow and hot pixels, were saved in a file with {\tt acis\_build\_badpix}. The event file information was then updated with {\tt acis\_process\_events}. A histogram of the duration vs. pointing and roll offsets was created with {\tt asphist}, and this was later used to create the auxiliary response files. Periods of anomalously high or low counts between 2.3--7.3 keV were removed with {\tt lc\_clean}. Point sources were excluded manually from a 0.5--4.0 keV image with circular regions a minimum of 25$\arcsec$ in radius created in SAOImageDS9\footnote{\url{https://sites.google.com/cfa.harvard.edu/saoimageds9}} (see Fig. \ref{chandra_images}). Any visible point sources not excluded were determined to be negligible. The final GTI for observation 1669 is $\sim$42 ks for the front-illuminated chips and $\sim$40 for the back-illuminated chips, and the final GTI for observation 6102 is $\sim$7 ks.
ACIS blank sky backgrounds that match the observation were obtained using {\tt acis\_backgrnd\_lookup}. These backgrounds were then tailored to match the data by also being filtered for bad pixels, aligned with the observation, and scaled by the 9.5--12.0 keV counts in the observation event file. A script written by Maxim Markevitch, called {\tt make\_readout\_bg}\footnote{\url{ https://cxc.cfa.harvard.edu/contrib/maxim/make_readout_bg}}, was used to create a model background file to correct for other ACIS readout artifacts. The raw and cleaned \chandra\ images are shown in Fig. \ref{chandra_images}.

The spectrum was extracted with the {\tt CHAV}\footnote{\url{http://hea-www.harvard.edu/~alexey/CHAV/}} tool {\tt runextrspec}. The data reduction, background creation, and spectral extraction steps we followed are described in more detail by \citet{wang2016}, however, we fit the spectra jointly in {\tt Xspec} rather than combining them as done by \citet{wang2016}.

\subsection{NuSTAR}
The two \nustar\ observations used in this work include both FPMA and FPMB spectra (Table \ref{data}). A user-defined GTI was created by removing flares with {\tt lcfilter}\footnote{\url{https://github.com/danielrwik/reduc}}. This command creates light curves from the A and B modules separately, binned by 100 seconds. Bins with count rates greater than the local distribution were identified manually and excluded, resulting in a $\sim$2--3$\sigma$ cut. The final GTIs were $\sim$98 and $\sim$122 ks for 70660002002 and 70660002004, respectively. This process is described in more detail by \citet{rojas2021}.

\begin{figure}[h]
 \centering
 \includegraphics[scale=0.625]{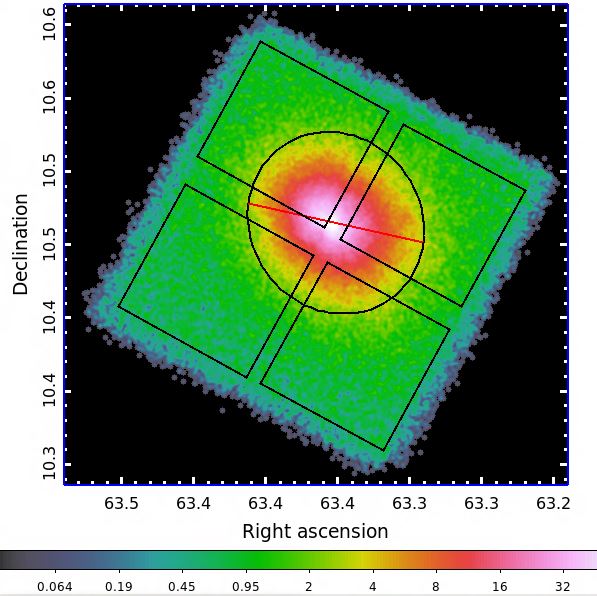}
 \caption{The raw 4-25 keV counts image of \nustar\ observation 70660002004. The background was extracted from the rectangular regions via {\tt nuproducts} and fit with the {\tt nuskybgd} code.}
 \label{raw_nustar}
\end{figure}

\begin{figure}[h]
 \centering
 \includegraphics[scale=0.3]{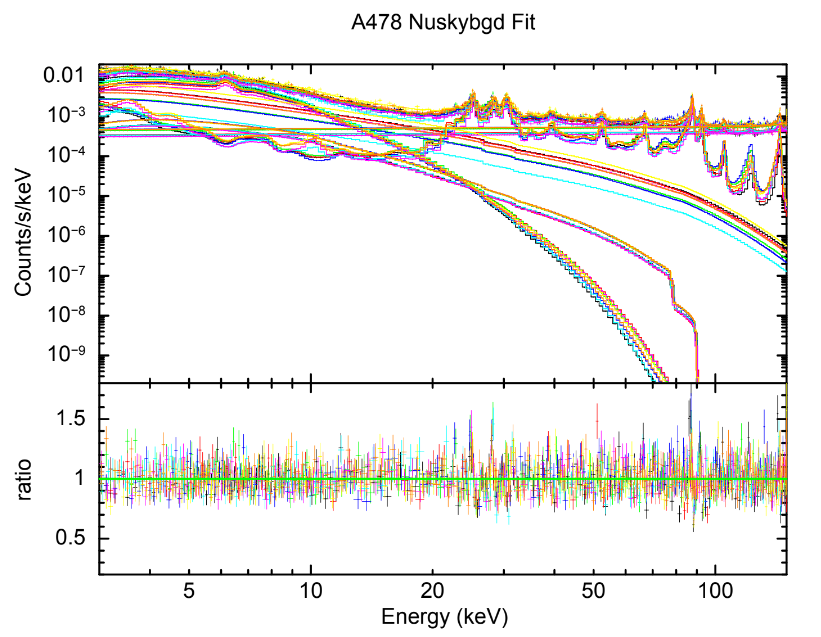}
 \caption{The FPMA and FPMB background fit of \nustar\ observation 70660002004; this is a typical fit, and the results of the background fit for observation 70660002002 are similar. This fit was done with {\tt nuskybgd}.}
 \label{bgfit}
\end{figure}

\begin{figure}[h]
 \centering
 \includegraphics[scale=0.24]{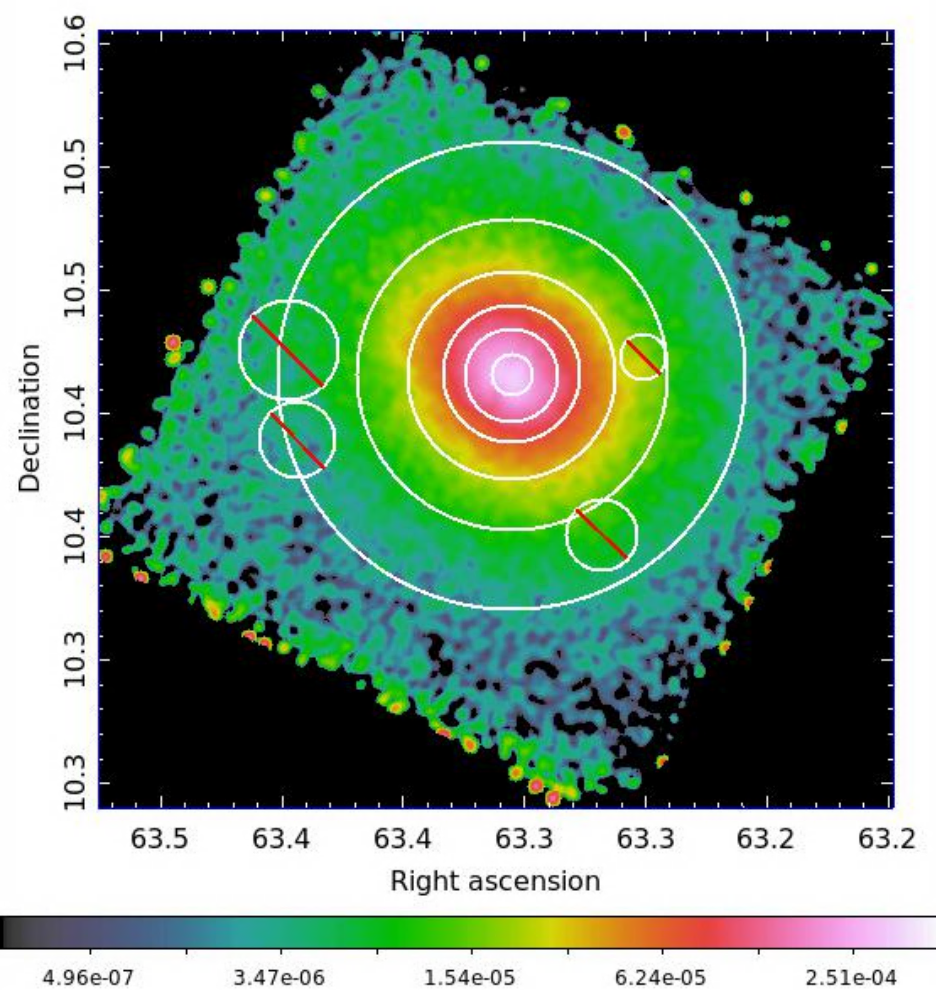}
 \caption{The exposure corrected, background subtracted 4--25 keV image of \nustar\ observation 70660002004. The spectrum was extracted from each of the annuli shown, excluding the point sources. The annuli have the same radii as shown in Figure \ref{chandra_images}, though the outermost annulus is not included. These spectra were fit to get the observed temperature profile.}
 \label{nustar_ann}
\end{figure}

For the \nustar\ data reduction, we use HEASoft version 6.28, NuSTARDAS\footnote{\url{https://heasarc.gsfc.nasa.gov/docs/nustar/analysis/}} version 2.0.0, and CALDB index version 20200912.
The background spectra were extracted via {\tt nuproducts} from square regions a little smaller than the chips, with an elliptical exclusion region to remove cluster emission (see Fig. \ref{raw_nustar}). These background spectra were fit using {\tt nuskybgd}\footnote{\url{https://github.com/NuSTAR/nuskybgd}}; this code models the solar, focused cosmic x-ray, aperture, and internal background components, as well as an {\tt APEC} component for the residual cluster emission \citep{wik2014}. The {\tt APEC} model is sufficient to account for all cluster emission not excluded by the elliptical exclusion region. The temperature, abundance, and normalization parameters of the {\tt APEC} model were left free to vary but tied across FPMA and FPMB. The redshift was fixed at 0.0856 \citep{redshift}. A constant component was included to account for the cross-calibration of the A and B modules, where the parameter value was fixed at 1 for FPMA and free to vary for FPMB. The final background fits for observation 70660002004 are shown in Fig. \ref{bgfit}. Following the background fit, the spectra were extracted from the annuli via using {\tt nuproducts} with {\tt bkgextract=no} because we use the backgrounds fit with {\tt nuskybgd}. The point source regions excluded were the same as for \chandra.\ The resulting background subtracted, exposure-corrected image can be seen in Fig. \ref{nustar_ann}.
%\todo{It may be worth going into a bit more detail on how the cluster component was modeled (free kT/abund/etc in the regions?  A \& B params tied at all?  The 2 outer annuli at least cover the background regions, so any degeneracy or bias could affect those fits.}
%\todo{You definitely need to expand on these descriptions and focus more on what is being done as opposed to how (or what tools) are being used. With the nustar analysis for example, you want to say how flares were identified--criteria used, essentially a rough, by eye sigma cut equivalent to $\sim$2-3$\sigma$--and how much the exposure time is reduced by.  Then say nuskybgd is used to characterize the bgd and create bgd spectra, citing my paper for details.  At a minimum, you should describe the regions used to characterize the bgd, including exclusion region, etc.  Look at Randall's A2163 paper to see an example.  He includes light curves and images and spectra of this part as figures -- this isn't strictly necessary, but can be easier than trying to describe things that are immediately clear visually.}

%\todo{In terms of figures, you definitely should show raw and bgd-sub, exp-corr images of both the chandra and nustar data, ideally showing the annular regions on the images as well.}

\subsection{XMM-Newton}
The center of A478 was observed by \xmm\ in 2002 for 126.6~ks \citep[OBSID 0109880101; first reported in][]{pointe04}.
A shorter offset pointing also covers part of the cluster, but we only analyze data from the longer central pointing.
Images and spectra are extracted following the Extended Source Analysis Software ({\textit XMM}--ESAS)\footnote{\url{https://heasarc.gsfc.nasa.gov/docs/xmm/xmmhp\_xmmesas.html}} package, included as part of SAS version 20.0.0, with a calibration analysis date of 2022-05-13. We used the standard data filtering and processing techniques as part of the analysis procedure \citep[e.g.,][]{snowden_esas}.
%\todo{If you know it, you can mention for XMM the date of the calibration files that were used (e.g., the "analysisdate" parameter in the cifbuild task)} \todo{caldb version?}
The clean %flare-removed?
exposure times for the three EPIC instruments amounted to 55.7~ks, 94.2~ks, and 62.6~ks for Metal Oxide Semi-conductors 1 and 2 (MOS1 and 2) and PN, respectively.
In addition to the annular regions described above for \chandra\, \nustar\, and \xmm\ spectra extraction, spectra from an outer annular region extending out to 13\arcmin\ were also extracted in order to better constrain foreground Galactic emission and absorption, which were modeled and simultaneously fit with the annular regions following \citet{snowden_esas}.

\begin{figure*}[ht]
 \centering
 \includegraphics[scale=0.6]{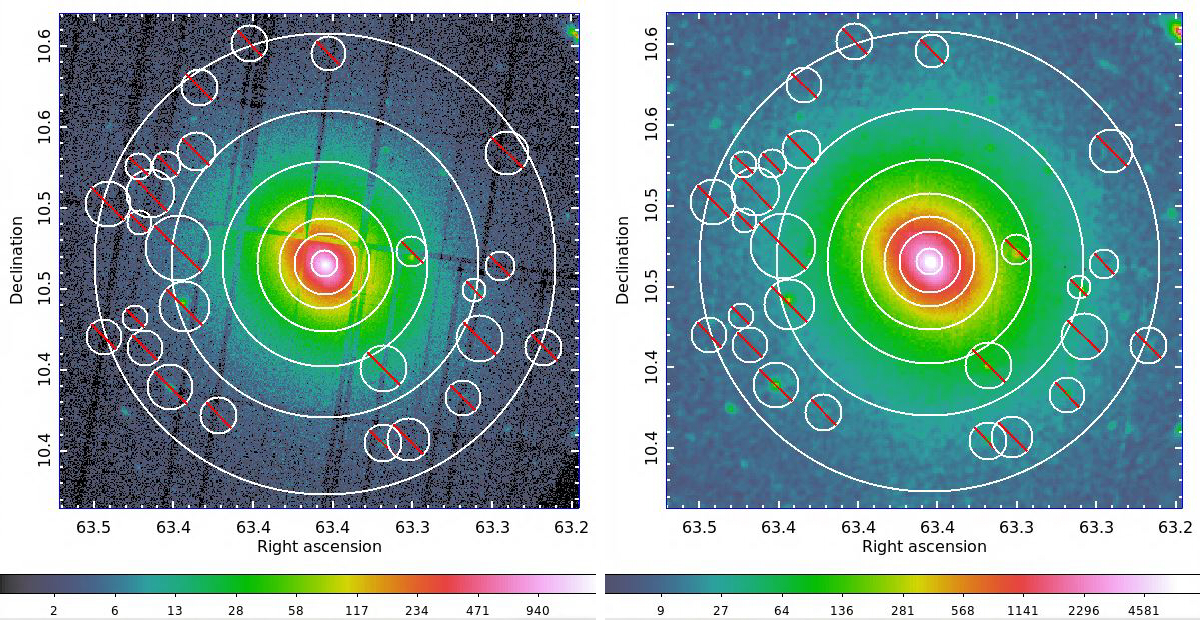}
 \caption{The raw (Left) and background subtracted, exposure corrected (Right) 0.4--7.2 keV counts images of the combined MOS1/MOS2/PN \xmm\ observation. The spectrum was extracted from each of the annuli shown;  these spectra were fit to get the observed temperature profile. The excluded regions are the same as used for \chandra\ and \nustar.}
 \label{xmm_images}
\end{figure*}

We also make use of two corrections to the effective area calibration, which are applied in the call to the SAS task {\tt arfgen} by setting two keywords: {\tt applyabsfluxcorr=yes}\footnote{\url{https://xmmweb.esac.esa.int/docs/documents/CAL-TN-0230-1-3.pdf}} and {\tt applyxcaladjusment=yes}\footnote{\url{https://xmmweb.esac.esa.int/docs/documents/CAL-TN-0018.pdf}}.
The former correction increases the $E>4$~keV EPIC effective area by a few percent, which brings the observed spectral indices of bright point sources observed simultaneously by \xmm\ and \nustar\ into agreement; although empirically determined only for EPIC-PN, the correction is applied to all three detectors for consistency. Despite the name, the {\tt applyabsfluxcorr} keyword does not make any corrections to the flux. The effect this correction has on the \xmm\ temperatures can be seen in Appendix \ref{xmm_fluxcorr}.
The latter correction further increases the hard band MOS effective areas to bring their measurements into better agreement with those of the PN.
These corrections bring \xmm\ detector spectra into a better agreement with each other and with \nustar\ spectra, although they do not necessarily ensure a {\it more accurate} calibration.
However, we note that simultaneous, absorbed single temperature fits to all annuli are better fit when these corrections are applied, and the residual soft proton contribution is also better constrained. The raw and cleaned \xmm\ images are shown in Fig. \ref{xmm_images}.

\section{Analysis and Results} \label{sec:analysis}

\subsection{Temperature and Density Profile Extraction}
%The annuli were chosen, following the method used by \citet{Vik05}, to start at a radius of $\sim30$ arcsec from the center of the cluster to avoid structure due to AGN activity. Each annulus had an outer radius of 1.5 times the inner radius. The inner two annuli were later combined, as the radius was too small to accurately correct for crosstalk due to NuSTAR's point spread function (PSF).

%\AT{We extracted spectra from both NuSTAR and Chandra data to draw radial temperature profiles.}
The \chandra,\ \nustar,\ and \xmm\ spectra were extracted from circular and annular regions to create a radial temperature profile. The chosen regions consist of a central circle of radius 30\arcsec\ and 5 concentric annuli starting at 30\arcsec, all centered at 63.3546$^{\circ}$, 10.4655$^{\circ}$.
The outer radius of each annulus was 1.5x the inner radius, though the inner two annuli later had to be combined to ensure each region was large enough to accurately correct for crosstalk due to \nustar's\ point spread function (PSF); the code used to correct for crosstalk, described in section \ref{psf}, requires regions to have a radius $\geq 30\arcsec$. This logarithmic spacing was chosen to maintain high signal to noise. Because they have a larger field of view than \nustar,\ one extra annulus was included for both \chandra\ and \xmm\ (see Figures \ref{chandra_images} and \ref{xmm_images}).

All of the spectra were fit using {\tt Xspec}\footnote{\url{https://heasarc.gsfc.nasa.gov/docs/xanadu/xspec/}}. The \chandra\ and \nustar\ spectra channels were grouped with grppha\footnote{\url{https://heasarc.gsfc.nasa.gov/docs/journal/grppha4.html}} to a minimum of 3 counts, and the C statistic was used to fit them. While the C statistic is more appropriate, it is also more computationally exspensive. Because of this, the \xmm\ spectra were grouped to a minimum of 30 counts, and the chi-square statistic was used. The choice of statistic has a negligible effect on the resulting best-fit temperatures
according to tests, and thus this choice should not affect any results.

The {\tt APEC} and {\tt TBABS} models were used to fit all spectra. The \chandra\ spectra of both obsids were jointly fit between 0.8--9 keV, and all parameters besides the {\tt APEC} norm were tied. The \nustar\ FPMA/B spectra of both obsids were fit between 3--15 keV, and the \xmm\ MOS1/2 and PN spectra were fit between 0.4--11~keV. 

The Wilms \citep[{\tt XSpec} table {\tt wilm};][]{wilms} abundances were used for all fits; using the Anders and Grevesse \citep[{\tt angr};][]{angr} abundances instead resulted in temperature differences of around $1\%$ at most for \nustar\ and around $5\%$ for both \chandra\ and \xmm. The Wilms table was used because it agrees better with HI measurements, as well as being generally better for fitting Galactic absorption (\citet{wilms}; \citet{willingale2013}). While \nustar\ abundances are estimated based on the Fe complex alone, \chandra\ and \xmm\ measurements are also impacted by unresolved lines at lower energies. However, some of the flux at these lower energies is also affected by the amount of line-of-sight absorption ($N_H$); if the $N_H$ is over- or underestimated, it will affect the abundance estimates as well. This degeneracy can cause differences in both abundance and $N_H$ measurements between the different telescopes and can be particularly sensitive to the accuracy of the calibration at low energies. 

In this work, the \xmm\ abundances are, excluding the outermost region, an average of $\sim$10\% lower than the \chandra\ abundances. In the outermost region, the \xmm\ abundance is $\sim$66\% larger than the \chandra\ abundance. The average difference between the \chandra\ and \xmm\ abundances is $\sim$0.1 $Z_{\odot}$. Changing the \chandra\ abundances by this amount results in a difference of $\sim$0.02 keV in the center and $\sim$0.13 keV in the outskirts. These differences are smaller than the $1\sigma$ confidence on the temperatures. Similarly, changing the \xmm\ abundances by $\sim$0.1 $Z_{\odot}$ results in temperature differences of $\sim$0.01 keV in the center, and $\sim$0.10 keV in the outskirts, which are on the order of the $1\sigma$ errors (see Table \ref{temps}). The \nustar\ abundances are an average of $\sim$38\% lower than the \chandra\ abundances. This is a difference of $\sim$0.2 $Z_{\odot}$. Changing the \nustar\ abundances by this amount results in a difference of $\sim$0.05 keV in the center and $\sim$0.1 keV in the outskirts, which is a little larger than the $1\sigma$ errors.

The $N_H$ along the line of sight of A478 changes with cluster radius and is larger than the radio value of $1.5\times10^{21}$ from the HI4PI survey \citep{HI4PI}. \citet{Vik05} find that the best-fit \chandra\ $N_H$ changes linearly with radius from $3.09\pm0.09\times10^{21}$ cm$^{-2}$ at the center to $2.70\pm0.06\times10^{21}$ cm$^{-2}$ at $4\arcmin$. \citet{pointe04} find that the best-fit \xmm\ $N_H$ also changes with radius from $3.00\pm0.10\times10^{21}$ cm$^{-2}$ at $0.14\arcmin$ to $2.41\pm0.08\times10^{21}$ cm$^{-2}$ at $4.54\arcmin$. These profiles agree in the center but progress to a difference of $\sim$11$\%$ around $4\arcmin$. In this work, $N_H$ was left free for each region individually for both \chandra\ and \xmm.\ This results in the \xmm\ $N_H$ values being an average of $\sim$13\% lower, with the maximum difference in the outermost region, where the \xmm\ $N_H$ is $\sim$26\% lower. This is an average difference between the \chandra\ and \xmm\ $N_H$ of $\sim5\times10^{20}$ cm$^{-2}$ (see Table \ref{temps}). Changing the \chandra\ or \xmm\ $N_H$ values by this amount results in a temperature difference of $\sim$0.4 keV in the center and up to $\sim$1 keV in the outskirts for both telescopes. Setting $N_H$ at the radio value found by \citet{HI4PI} more than doubles the temperatures for both telescopes as well. The temperature profile that results from fixing the \chandra\ $N_H$ to the \xmm\ best-fit values, as well as the \xmm\ spectra with $N_H$ fixed to the \chandra\ best fit values can be seen in Appendix \ref{abs}. \nustar\ is less sensitive to absorption \citep{tumer2023}; in the central region, the inclusion of $N_H$ fixed to the best-fit \chandra\ value decreases the temperature by $\sim$1\% compared to a fit without it.

A detailed investigation into these differences is beyond the scope of this work, which considers absorption and abundance to be nuisance parameters even though they could bias temperature estimates from \chandra\ and \xmm.

The {\tt APEC} norms also differ between telescopes. Since the norm depends on the electron and proton densities, differences in the norms imply differences in flux. The \nustar\ fluxes are an average of $\sim$9\% and $\sim$4\% lower than the \chandra\ ACIS-S and ACIS-I fluxes respectively, excluding \nustar’s\ outermost region (\chandra’s\ second to last region), where the \nustar\ flux is $\sim$40\% higher than the ACIS-S flux, and $\sim$3\% higher than the ACIS-I flux. The two outermost \chandra\ regions contain very little of the ACIS-S observation, so this is expected.
The \xmm\ (MOS1) fluxes are an average of $\sim$24\% lower than the ACIS-S \chandra\ fluxes, excluding the two outermost regions, where the \xmm\ fluxes are $\sim$13\% and $\sim$65\% larger, respectively. The \xmm\ fluxes are an average of $\sim$25\% lower than the ACIS-I \chandra\ fluxes. The \xmm\ fluxes are also an average of $\sim$18\% lower than the NuSTAR fluxes. Thus, there is tension not only in the temperatures but in the fluxes as well.

The redshift for \nustar\ was fixed at $0.0856$, and a gain offset was added and free to fit. The resulting gain is $-0.102^{+7e-6}_{-0.001}$~keV.
%drw add
This adjustment is needed to eliminate systematic residuals in the Fe K complex while obtaining the known redshift of the cluster; similar gain adjustments of $\sim$0.1~keV are required when fitting other clusters, suggesting a small miscalibration at lower energies \citep{rojas2021}.
The \chandra\ redshift was free to vary in each region to account for any differences in gain between them due to spanning across both the ACIS-S and ACIS-I observations. The resulting best-fit values are an average of $\sim$4$\%$ higher than the accepted value of 0.0856 \citep{redshift}, excluding the second to last region, which has a best-fit value $\sim$10$\%$ lower. Fixing the redshift in this region at 0.0856 increased the temperature by 0.02 keV, or $\sim$0.3$\%$, which is less than the 1$\sigma$ temperature uncertainty of -0.31,+0.17 keV. The redshift was free to vary but tied across regions for \xmm\ to account for gain. The resulting best-fit value is $\sim$5$\%$ lower than the accepted value. %\todo{Garrit: You should have a good idea about the gain through the fluorescent lines.}

%{\tt mkprof}, a tool in Alexey Vikhlinin's {\tt ZHTOOLS}\footnote{\url{https://github.com/avikhlinin/zhtools-mac}}. 

The projected surface brightness was extracted by calculating the count mean in the background subtracted, exposure corrected (flux) mosaic \chandra\ image in logarithmic radial bins. The image is masked with the excluded point source regions. The uncertainty is calculated assuming Poisson statistics. This surface brightness profile was used for determining the \chandra,\ \nustar,\ and \xmm\ mass estimates because \chandra\ has higher spatial resolution than both \nustar\ and \xmm. This profile was fit using a double beta model and emissivity table, as discussed in section \ref{fitting}.

%using the raw counts and exposure map images (also masked with the excluded point sources):}

%\todo{Put in an equation: square root of the raw count profile times N over the exposure map times N (where N is the number of points?)}

%to create that radial profile, you summed over regions of the raw counts, simulated background counts, and exposure map images separately, then computed the flux and uncertainty in each radial bin assuming poisson statistics.

%\todo{Gerrit: Do you fit this with a (double) beta model? How is the background taken into account?} 

%\vfill\null

\subsection{Crosstalk Correction}
\label{psf}
%\todo{Ben: You should discuss the fact that the XMM PSF is not modeled.}
\nustar\ has a larger PSF than both \chandra\ and \xmm.\ Photons from different parts of the gas are mixed and need to be disentangled. To correct for this crosstalk between the annular regions, {\tt nucrossarf}\footnote{\url{https://github.com/danielrwik/nucrossarf}} was used; this code creates new response matrix files (RMF), ancillary response files (ARF), and background files based on the shape of \nustar's\ PSF. These new files were used to jointly fit the annular spectra in {\tt Xspec}. For the source image, or the image that gives {\tt nucrossarf} the actual distribution of photons, we used a point-source masked and filled \xmm\ image in the 2.35--7.2~keV band because the \nustar\ image has already been smoothed by the PSF. We do not use the \chandra\ image because, while \chandra\ has the greatest spatial resolution, the \chandra\ image has more artifacts from combining ACIS-S and ACIS-I observations. Additionally, \xmm\ has a greater effective area in the hard band, thus it more closely resembles \nustar\ images. How the {\tt nucrossarf} code disentangles the crosstalk is explained in detail by \citet{tumer2023}.

\xmm's\ PSF is also larger than \chandra's.\ However, with it being smaller than \nustar's\ PSF, and with the regions being as large as they are, we do not correct \xmm's\ PSF.

%\AT{Here, give a reference to the github page as a footnote, just like you did for acisgenie. By the time you submit, hopefully my paper will be published and you can refer to it for crossarf and maybe to Randall's paper when it's on arxiv}

\subsection{\nustar\ Temperature Map}

A temperature map of the \nustar\ observation 70660002002 was made by first separating the spectrum into narrow energy bands of 3--5 keV, 6--10 keV, and 10--20 keV. A circle was drawn at every 5th pixel, which had a minimum radius of 5 pixels. Then the radius of the circle was increased until 1000 counts were reached to ensure $<$10\% precision measurements. This was done for the raw count, background, and exposure images for all 3 bands.

\begin{figure}[h]
 \centering
 \includegraphics[scale=0.24]{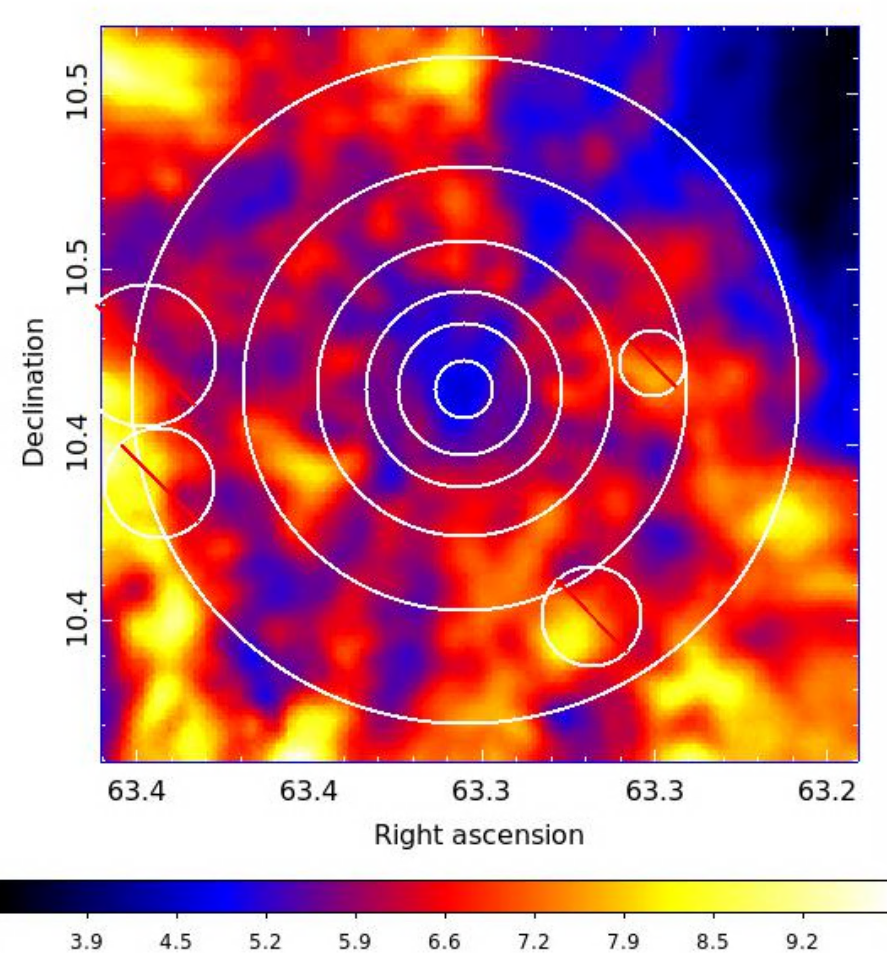}
 \caption{The temperature map of \nustar\ observation 70660002002. The annular regions and excluded point sources used in the spectral analysis are shown; the points were not excluded for the creation of the temperature map. The very cold area in the upper right corner is due to the map going off the \nustar\ field of view.}
 \label{tmap}
\end{figure}

The exposure corrected, background subtracted counts in each of these bands together form a rough spectrum that is fit in {\tt Xspec} for each circle. The center pixel of each circle is then assigned the resulting best-fit temperature. Then the temperature is interpolated between each of the central pixels in order to make an image. The \nustar\ temperature map made with this method can be seen in Fig. \ref{tmap}. 

The temperature map shows that the core is cooler, and temperature generally increases with radius. Multiple hot spots can be seen within the annular regions. Those that do not appear to correspond to point sources could be areas of hot gas in the cluster, and thus were not excluded in the spectral extraction. 

To derive a radial temperature profile from the temperature map, we create concentric annular regions and find the average or median temperature from all the measurements inside the region.  This profile can then be compared to the temperature profile derived by directly fitting spectra in annular regions using {\tt nuproducts} response files (Fig. \ref{temp_prof}), to evaluate the accuracy of the temperatures in the map.  We find that the profile is in good agreement with the {\tt nuproducts}-derived profile, except that the profile from the map is systematically lower by $\sim$0.25 keV.  This indicates a small systematic difference between how the temperatures are estimated in the two cases---likely resulting from the much more coarse-grained nature of the temperature map fits---but confirms the general consistency of the temperature map measurements and that from direct spectral fitting.  However, in both cases the measured temperatures suffer from spatial mixing due to the PSF; the {\tt nucrossarf}-derived temperatures are the best temperature estimates, since they account for this effect, and are what we use for the \nustar\ temperature profile and mass estimates.

\subsection{Comparing Temperature Profiles}\label{sec:temps}

\citet{Vik05} compare the observed temperature profile from \chandra\ and \xmm\ of A478, in addition to several other clusters. For most clusters, the \xmm\ temperature profile agrees with the \chandra\ profile after a +10\% renormalization to account for cross-calibration. For A478 however, the profiles are significantly different within the central 4\arcmin; the \xmm\ profile increases gradually before becoming relatively constant, whereas the \chandra\ profile increases more rapidly, peaks, and begins to decrease. This is attributed to an incomplete correction of \xmm’s\ PSF and the complex temperature structure in the cluster center.

\begin{figure}[h]
 \centering
 \includegraphics[scale=0.35]{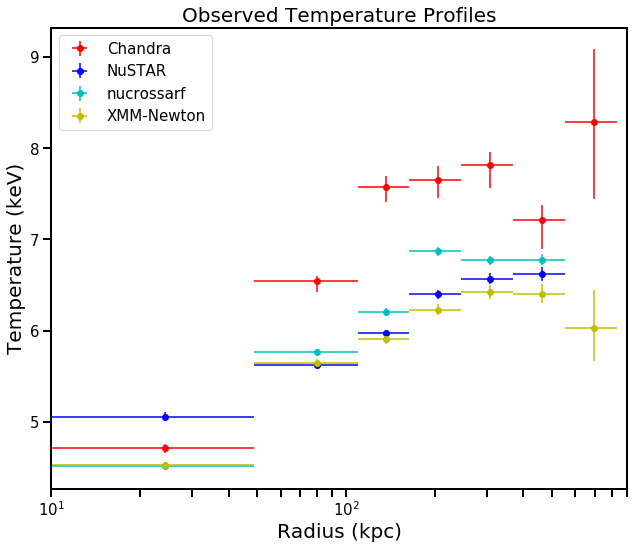}
 \caption{The observed temperature profiles from \chandra,\ \nustar,\ and \xmm.\ The initial \nustar\ fit is in blue and the fit after running {\tt nucrossarf} is in cyan.}
 \label{temp_prof}
\end{figure}

We do not correct for \xmm's\ PSF in this work, but \nustar's\ PSF has been corrected with {\tt nucrossarf}. The scattering from the PSF smoothed out the profile, making the inner region appear hotter and the outer regions appear cooler. After correction, the inner region is much cooler, and the outer regions are hotter. While this creates a more rapid increase to the peak, the {\tt nucrossarf} profile is still in better agreement with \xmm\ than \chandra\ (see Fig. \ref{temp_prof} and Table \ref{temps}). The \nustar\ temperatures are around 11\% lower than \chandra\ on average, with the largest difference in the 3rd region, where \nustar\ is 18\% cooler than \chandra.\ The largest difference between the \nustar\ and \xmm\ profiles is in the 4th region, where the \xmm\ temperature is $\sim$10\% lower than the \nustar\ temperature. The \xmm\ temperatures are lower than the \nustar\ temperatures by $\sim$5\% on average. For a comparison to temperature profiles found in other works, see Appendix \ref{other_temps}.

\begin{table*}
\centering
\begin{tabular}{ccccccc} 
\multicolumn{7}{c}{\large{\textbf{{\tt Xspec} Fit Parameters}}}\\
\hline
\hline
 &Radii& {\it kT} &{\it Z} & {\it N$_{H}$} & {\it norm} & \\
 & (arcsec) & (keV) & ($Z_{\odot}$) & (10$^{22}$~cm$^{-2}$) & ($10^{-2}$) & {\it z} \\
\hline
\\[-0.95em]
{\multirow{7}{*}{\rotatebox[origin=c]{90}{\chandra\ }}}&0$-$30 & $4.71^{+0.05}_{-0.05}$ & $0.71^{+0.02}_{-0.01}$ & $0.457^{+0.005}_{-0.003}$ & $1.641^{+0.005}_{-0.008}$ & $0.088^{+0.001}_{-0.001}$ \\
\\[-0.95em]
&30$-$68 & $6.54^{+0.06}_{-0.11}$ & $0.64^{+0.02}_{-0.03}$ & $0.441^{+0.005}_{-0.003}$ & $1.964^{+0.005}_{-0.009}$ & $0.086^{+0.001}_{-0.001}$ \\
\\[-0.95em]
&68$-$101 & $7.57^{+0.13}_{-0.16}$ & $0.54^{+0.03}_{-0.04}$ & $0.426^{+0.006}_{-0.004}$ & $1.218^{+0.006}_{-0.008}$ & $0.086^{+0.001}_{-0.001}$ \\
\\[-0.95em]
&101$-$152 & $7.65^{+0.15}_{-0.19}$ & $0.65^{+0.02}_{-0.08}$ & $0.409^{+0.006}_{-0.004}$ & $1.136^{+0.006}_{-0.005}$ & $0.084^{+0.001}_{-0.001}$ \\
\\[-0.95em]
&152$-$228 & $7.82^{+0.14}_{-0.26}$ & $0.43^{+0.03}_{-0.06}$ & $0.410^{+0.005}_{-0.005}$ & $1.042^{+0.004}_{-0.006}$ & $0.086^{+0.002}_{-0.002}$ \\
\\[-0.95em]
&228$-$342 & $7.21^{+0.17}_{-0.31}$ & $0.48^{+0.06}_{-0.05}$ & $0.375^{+0.010}_{-0.007}$ & $0.497^{+0.003}_{-0.006}$ & $0.077^{+0.006}_{-0.002}$ \\
\\[-0.95em]
&342$-$513 & $8.29^{+0.80}_{-0.85}$ & $0.41^{+0.21}_{-0.22}$ & $0.275^{+0.034}_{-0.022}$ & $0.124^{+0.005}_{-0.005}$ & $0.087^{+0.008}_{-0.010}$\\
\\[-0.95em]
\hline

{\multirow{7}{*}{\rotatebox[origin=c]{90}{{\tt nucrossarf}}}}&0$-$30 & $4.52^{+0.02}_{-0.02}$ & $0.43^{+0.01}_{-0.01}$ & 0.457$^{*}$ & $1.648^{+0.010}_{-0.010}$ & 0.0856$^{\dagger}$ 
\\
\\[-0.95em]
& 30$-$68 & $5.77^{+0.02}_{-0.02}$ & $0.48^{+0.01}_{-0.01}$ & 0.441$^{*}$ & $2.192^{+0.008}_{-0.009}$ & - \\
\\[-0.95em]
&68$-$101 & $6.21^{+0.04}_{-0.04}$ & $0.29^{+0.02}_{-0.02}$ & 0.426$^{*}$ & $1.336^{+0.008}_{-0.009}$ & - \\
\\[-0.95em]
&101$-$152 & $6.87^{+0.05}_{-0.05}$ & $0.44^{+0.02}_{-0.02}$ & 0.409$^{*}$ & $1.163^{+0.006}_{-0.007}$ & - \\
\\[-0.95em]
&152$-$228 & $6.77^{+0.05}_{-0.05}$ & $0.30^{+0.02}_{-0.02}$ & 0.410$^{*}$ & $1.014^{+0.006}_{-0.007}$ & - \\
\\[-0.95em]
&228$-$342 & $6.78^{+0.06}_{-0.06}$ & $0.22^{+0.02}_{-0.02}$ & 0.375$^{*}$ & $0.793^{+0.006}_{-0.006}$ & - \\
\\[-0.95em]
\hline

{\multirow{8}{*}{\rotatebox[origin=c]{90}{\xmm\ }}}&0$-$30 & $4.53^{+0.03}_{-0.03}$ & $0.68^{+0.01}_{-0.01}$ & $0.406^{+0.002}_{-0.002}$ & $1.262^{+0.003}_{-0.003}$ & $0.0816^{+0.0002}_{-0.0002}$ $^{\ddagger}$ 
\\
\\[-0.95em]
&30$-$68 & $5.65^{+0.04}_{-0.04}$ & $0.56^{+0.01}_{-0.01}$ & $0.390^{+0.002}_{-0.002}$ & $1.836^{+0.003}_{-0.003}$ & - 
\\
\\[-0.95em]
&68$-$101 & $5.91^{+0.06}_{-0.06}$ & $0.54^{+0.02}_{-0.02}$ & $0.383^{+0.003}_{-0.003}$ & $0.953^{+0.001}_{-0.001}$ & -
\\
\\[-0.95em]
&101$-$152 & $6.23^{+0.06}_{-0.06}$ & $0.44^{+0.02}_{-0.02}$ & $0.376^{+0.002}_{-0.002}$ & $1.030^{+0.004}_{-0.003}$ & -
\\
\\[-0.95em]
&152$-$228 & $6.43^{+0.07}_{-0.08}$ & $0.42^{+0.02}_{-0.02}$ & $0.355^{+0.002}_{-0.002}$ & $0.817^{+0.002}_{-0.002}$ & - 
\\
\\[-0.95em]
&228$-$342 & $6.40^{+0.11}_{-0.10}$ & $0.44^{+0.03}_{-0.03}$ & $0.327^{+0.003}_{-0.003}$ & $0.618^{+0.003}_{-0.003}$ & -
\\
\\[-0.95em]
&342$-$516 & $6.03^{+0.42}_{-0.36}$ & $0.68^{+0.09}_{-0.09}$ & $0.236^{+0.036}_{-0.034}$ & $0.222^{+0.002}_{-0.002}$ & - \\
\hline
\end{tabular}
\tablenotetext{*}{Fixed to the \chandra\ best-fit hydrogen column density values.}
\tablenotetext{\dagger}{Fixed to 0.0856 and tied across regions.}
\tablenotetext{\ddagger}{Free to fit and tied across regions.}
\tablecomments{The APEC norm is given by $\frac{10^{-14}}{4\pi\left[D_A\left(1+z\right)\right]}\int n_e n_H dV$ where $D_A$ is the angular diameter distance to the source, and $n_e$ and $n_H$ are the electron and proton densities, respectively.}
\caption{Spectral fit results for all annuli of \chandra,\ \nustar,\ and \xmm.\ }
\label{temps}
\end{table*}

\subsection{Temperature Weighting} \label{sec:weighting}
\begin{figure}[h]
 \centering
 \includegraphics[scale=0.4]{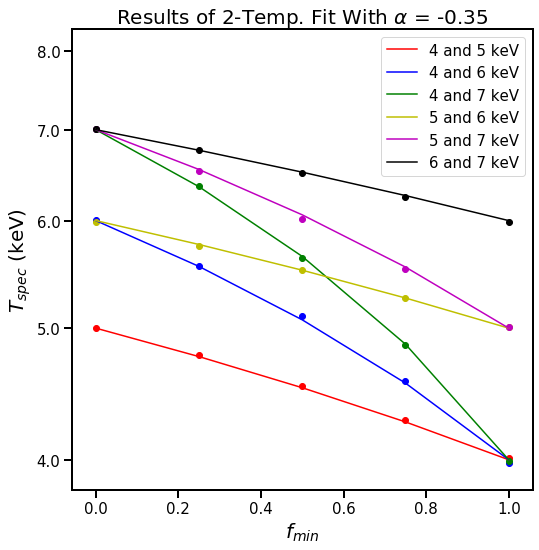}
 \caption{Results of the two temperature component simulations. The components were 4 \& 5 keV, 4 \& 6 keV, 4 \& 7 keV, 5 \& 6 keV, 5 \& 7 keV, 6 \& 7 keV, and the normalizations were $f_{min}$ and $(1-f_{min})$ respectively. The fits are the result of the weighting equation with $\alpha=-0.35$}.
 \label{weighting_sim}
\end{figure}

\begin{figure}[h]
 \centering
 \includegraphics[scale=0.55]{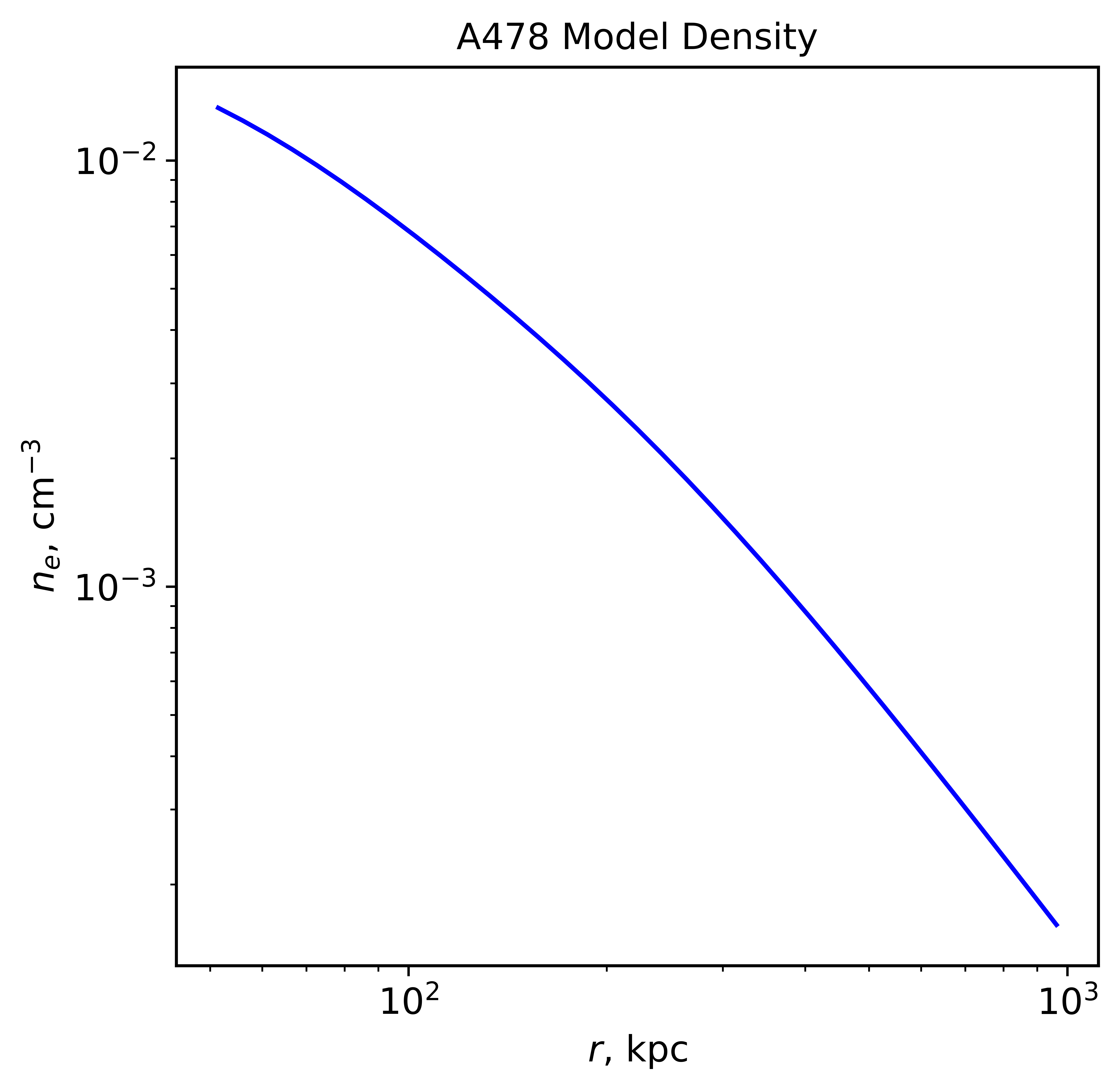}
 \caption{The double-$\beta$ \chandra\ density profile fit. The parameters for this fit are shown in Table \ref{pars2}.}
 \label{density}
\end{figure}

When converting to a projected temperature profile model, the 3D temperature profile model was weighted based on detector sensitivity. The following weighting formula was used \citep{Vik06_1}:
\begin{equation}
    \left<T\right>=\frac{\int_Vc(T)\rho^2T^{1-\alpha}dV}{\int_Vc(T) \rho^2T^{-\alpha}dV},
\end{equation}
where $c(T)$ is the detector sensitivity to bremsstrahlung radiation and $\rho$ is the density profile. For $\rho$ we use the \chandra\ density profile, as given in Fig. \ref{density}.
%This sensitivity is defined as the photon count rate for a spectrum with unit emission measure.
Simulations of \nustar\ spectra were done in {\tt Xspec} with temperatures 4 \& 5 keV, 4 \& 6 keV, 4 \& 7 keV, 5 \& 6 keV, 5 \& 7 keV, and 6 \& 7 keV, where the lower temperature has a norm of $f_{min}$, and the higher temperature has a norm of $1-f_{min}$. These were saved and then fit to single temperature models in the range 3--15 keV. The results of these fits are shown in Fig. \ref{weighting_sim}. The best-fit value of $\alpha$ for \nustar\ was found to be $-0.35$. The best value for \chandra\ and \xmm\ is found by \citet{Vik06_1} to be $0.75$. 
%\todo{Gerrit: I wonder if it would not just be easier to use Alexey's mixT code to derive these factors numerically including the line emission? Probably won't change much of your results though. But in case you want it for later, I can give you a half-python version of the code that I wrote.}

\subsection{Fitting the Temperature and Density Profiles}
\label{fitting}
\begin{figure}[h]
 \centering
 \includegraphics[scale=0.65]{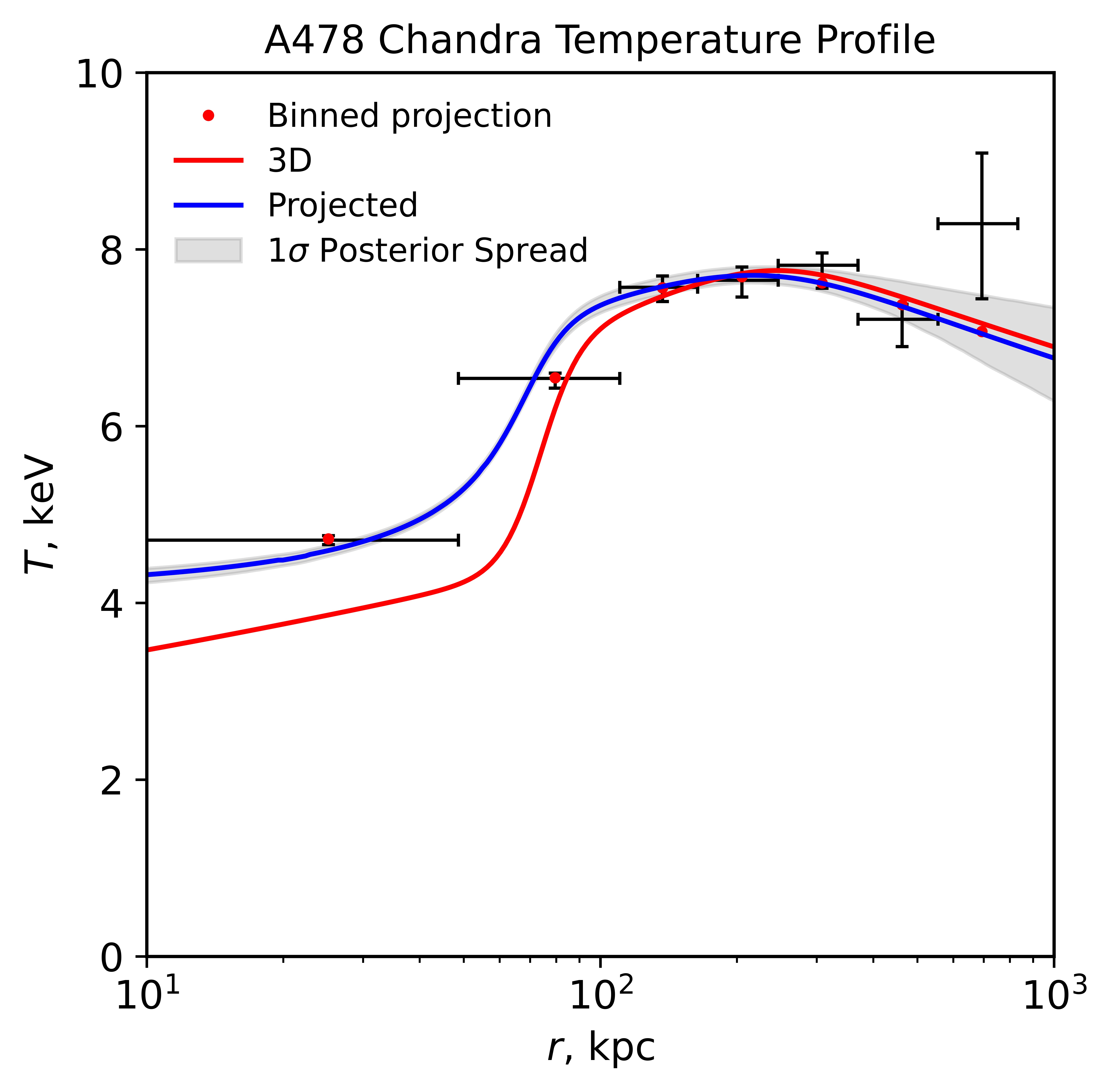}
 \caption{The projected (blue) and 3D (red) models of the \chandra\ temperature profile, fit with the $T_{cool}$ parameters free. The parameters for this fit are shown in Table \ref{pars}. The 1$\sigma$ posterior spread is found via Markov Chain Monte Carlo simulation as described in Section \ref{sec:MCMC}.}
 \label{chandra_model}
\end{figure}

\begin{figure}[h]
 \centering
 \includegraphics[scale=0.65]{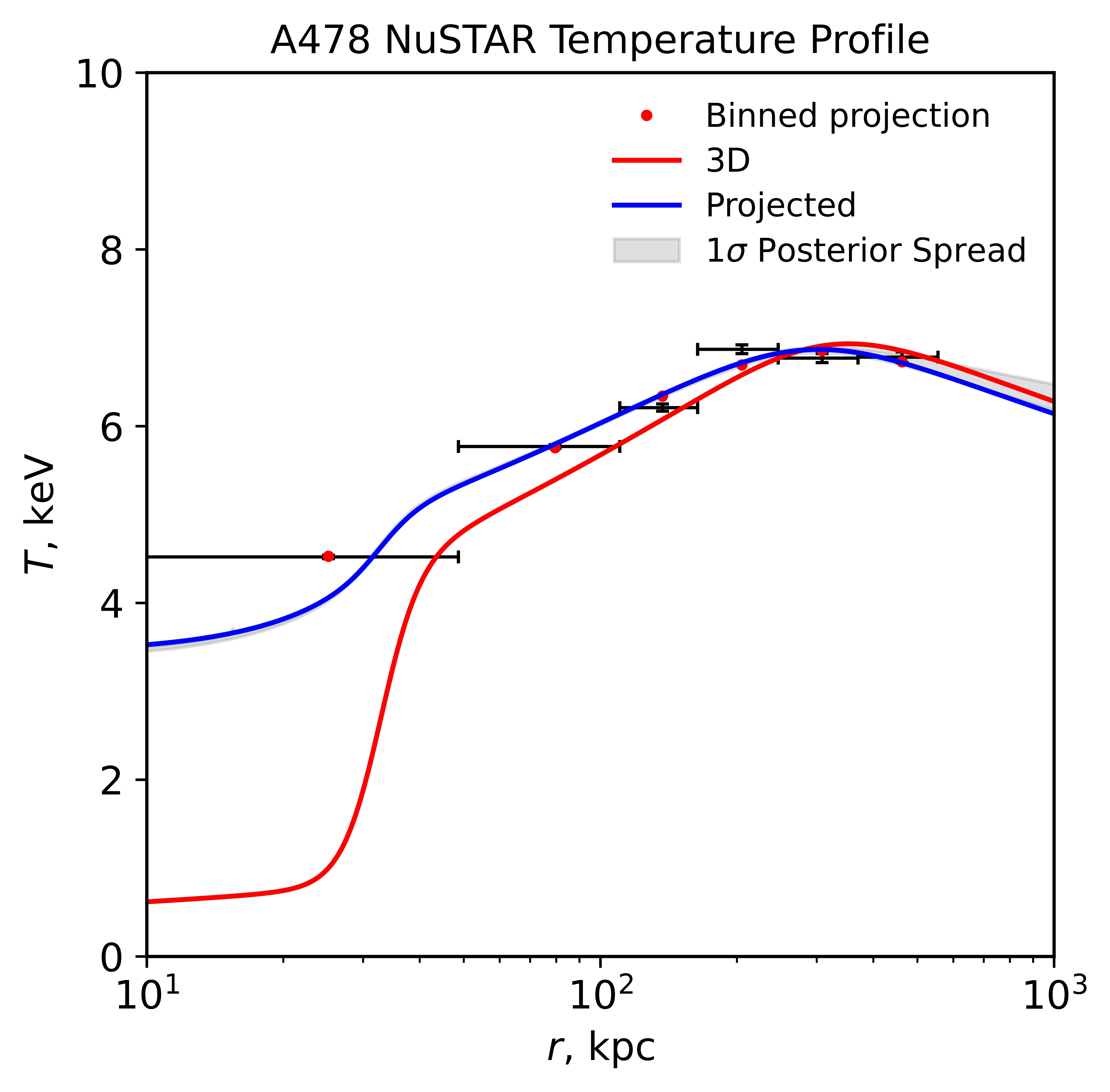}
 \caption{The projected (blue) and 3D (red) models of the \nustar\ temperature profile, fit with the $T_{cool}$ parameters free. The parameters for this fit are shown in Table \ref{pars}. The 1$\sigma$ posterior spread is found via Markov Chain Monte Carlo simulation as described in Section \ref{sec:MCMC}.}
 \label{nustar_model}
\end{figure}

\begin{figure}[h]
 \centering
 \includegraphics[scale=0.65]{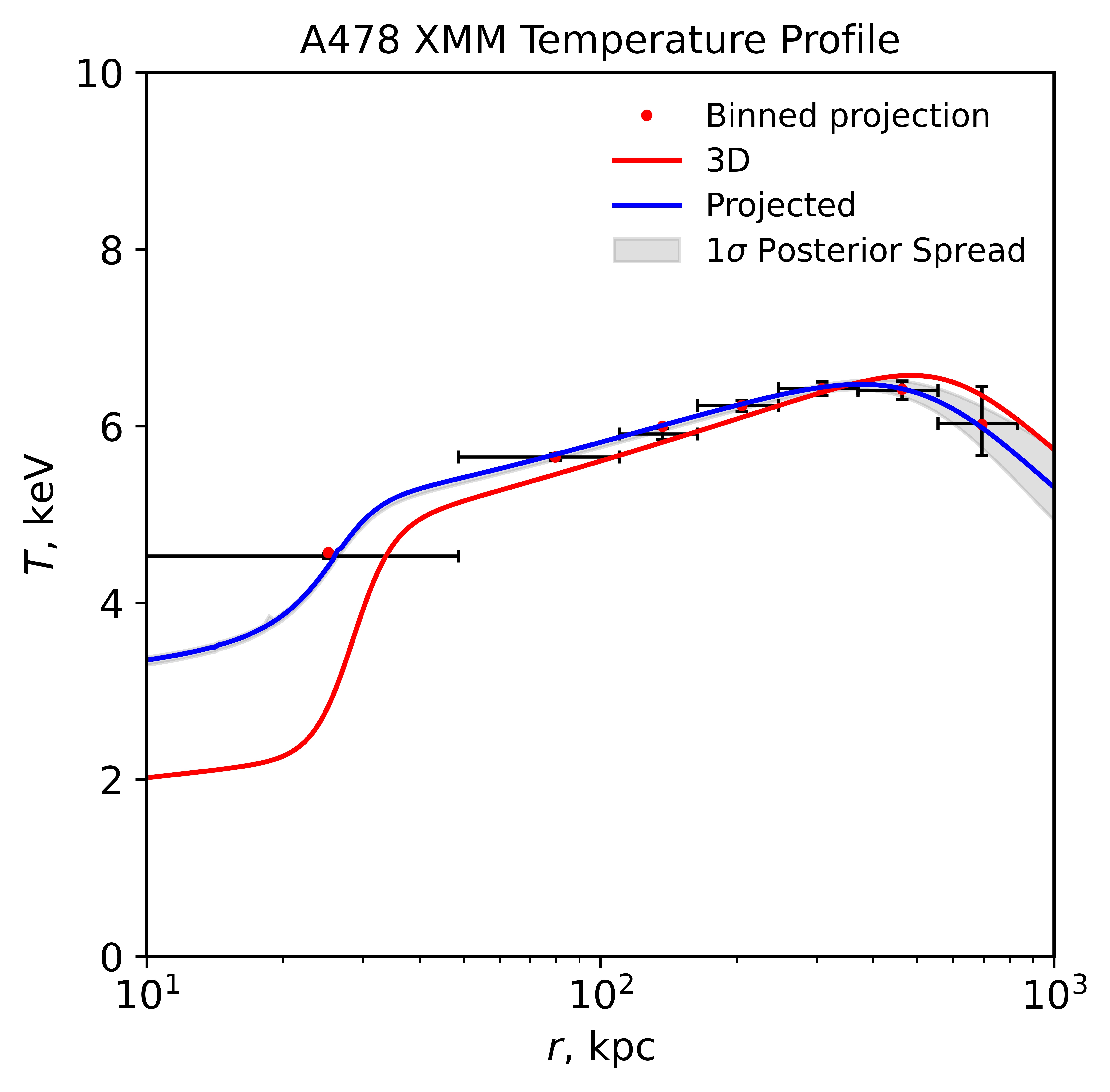}
 \caption{The projected (blue) and 3D (red) models of the \xmm\ temperature profile, fit with the $T_{cool}$ parameters free. The parameters for this fit are shown in Table \ref{pars}. The 1$\sigma$ posterior spread is found via Markov Chain Monte Carlo simulation as described in Section \ref{sec:MCMC}.}
 \label{xmm_model}
\end{figure}

The 3D temperature profile model used is that also used by \citet{vik06_2}, and is as follows:
\begin{equation}
    T_{3D}(r)=\frac{T_0 (r/r_t)^{-a}}{\left[1+(r/r_t)^b\right]^{c/b}}\times T_{cool},
\end{equation}
where the $T_{cool}$ component \citep{allen01} is described by \citet{vik06_2} as
\begin{equation}
    T_{cool}=\frac{(r/r_c)^{a_c}+T_{min}/T_0}{(r/r_c)^{a_c}+1}.
\end{equation}
Thus the model has nine parameters ($T_0$, $r_t$, $a$, $b$, $c$, $T_{min}$, $r_c$, $a_c$) to be fit. This 3D model cannot be fit directly to the observed temperature profile because the observed profile is actually a projection of the cluster temperature. After being weighted based on the weighting formula (see section \ref{sec:weighting}), the 3D temperature model is integrated along the line of sight to a truncation radius of 3500 kpc; this is the projected model (blue) in Figures \ref{chandra_model}, \ref{nustar_model}, and \ref{xmm_model}. The projected profile is then fit to the observed temperature profile and binned based on the inner and outer radii of the regions.

%\AT{explain what the parameters are}

The effect of $T_{cool}$ on the profile is focused mainly in the center, which has less of an effect on the final mass than the overall profile. The $T_{cool}$ parameters were fixed to the best-fit values found by \citet{vik06_2} for one fit, and left free to vary with the rest of the parameters for another. The rest of the parameters were left free for all fits, since the observed profiles of this work differ from the observed \chandra\ profile found by \citet{vik06_2} (see Section \ref{sec:temps}). The resulting mass at $r_{2500,Chandra}$ with fixed $T_{cool}$ parameters was $\sim$1\% lower for \chandra, $\sim$2\% lower for \nustar,\ and $\sim$7\% lower for \xmm.\

The 3D density model is a double-$\beta$ model \citep{Hudson10}:
\begin{equation}
    \label{density_equation}
    n(r)=\left[n_1^2(1+(r/r_1)^2)^{-3\beta_1}+n_2^2(1+(r/r_2)^2)^{-3\beta_2}\right]^{1/2}.
\end{equation}

An emissivity table was used to convert this density model to a surface brightness model. This table was constructed by first extracting the \chandra\ spectrum in a circular region with a 3$\arcmin$ radius centered on the cluster. After modeling the spectrum with the {\tt APEC} model in {\tt Xspec}, the {\tt APEC} norm was set to 1, and the model count rate was recorded for each point in a matrix of temperature and abundance values. Thus, the resulting density fit subsumed the {\tt APEC} norm and has to be normalized after the fit. Because the surface brightness profile comes from the \chandra\ mosaic image, including both the ACIS-S and ACIS-I observations, only the \chandra\ emissivity table was calculated. 
%\todo{
In future work, surface brightness and emissivity tables from \xmm\ and \nustar\ can be derived and used as well. Corrections for PSF cross-talk for both observatories must be applied to these processes. Because the focus of this work is on the impact of different temperature measurements between instruments, the use of only \chandra\ surface brightness and emissivity profiles should not significantly affect our results.
%For this work, we do not believe the use of only the \chandra\ surface brightness and emissivity profiles to (cause much uncertainty/error?)
%}

\begin{figure}[h]
 \centering
 \includegraphics[scale=0.6]{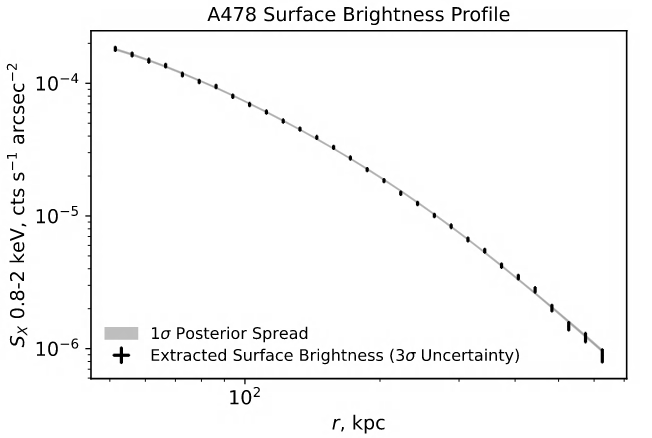}
 \caption{Surface brightness profile extracted from the mosaic \chandra\ image. The uncertainties on the extracted surface brightness points are given as $3\sigma$ to make them more visible. The 1$\sigma$ posterior spread shown is the \chandra\ model fit found via Markov Chain Monte Carlo simulation as described in Section \ref{sec:MCMC}. }
 \label{SB_profile}
\end{figure}

The emissivity table is converted to a function via interpolation. The function is then given the abundance and temperature model parameters to calculate the emissivity as a function of radius. This profile is then multiplied by the density model squared, then multiplied by 2 and integrated along the line of sight to a truncation radius of 3500 kpc; the resulting profile is fit to the observed surface brightness profile (see Fig. \ref{SB_profile}). Since the emissivity function depends on the temperature model fit, the density model fit will depend on the temperature model fit as well. Additionally, the temperature weighting, and thus the temperature model fit, depends on the density model fit. Thus, these models were iteratively fit with chi-squared minimization.

%\vfill\null

\subsection{Markov Chain Monte Carlo For Confidence Intervals} \label{sec:MCMC}
The python package {\tt emcee}\footnote{\url{https://emcee.readthedocs.io/en/v2.2.1/}} was used to run a Markov chain Monte Carlo (MCMC) simulation on the temperature and density profile parameters. MCMC is used to sample posterior probability distribution functions (pdf) by comparing pairs of points; this means that it is insensitive to pdf normalization, and does not need a full analytic description of the pdf \citep{Hogg18}.

The log-likelihood function, or the function that determines whether a set of parameters is accepted or not, comes from the residuals of both temperature and surface brightness models:
\begin{equation}
    -\frac{1}{2}\Sigma\left[\left(\frac{T-T_{model}}{T_{err}}\right)^2+\left(\frac{SB-SB_{model}}{SB_{err}}\right)^2\right]
\end{equation}

The code was given the best-fit parameters as a starting point, and then it was given a step size for each parameter, the number of walkers (50), and the number of iterations (1000). Because of the small effect of the $T_{cool}(r)$ term, the $T_{min}$, $r_c$, and $a_c$ parameters were fixed at the best-fit values for each telescope while running the chain. The acceptance fraction, on average, was $\sim$0.23; this indicates that the chosen step size was appropriate, being close to 0.234, the ideal for high dimensional models \citep{Hogg18}. Though the chain was not run for the recommended 50 times the integrated autocorrelation time, the walkers passed through the high probability areas of the parameter space many times, which indicates that the length of the chain was appropriate as well.

%The integrated autocorrelation time $(\tau_{int})$, or the number of iterations it takes for the result to be completely independent from the starting point, is a good test of how long the chain should be run. The sampling is most accurate when the chain is run for many times $\tau_{int}$. In this case, the chain was around 10 times longer than $\tau_{int}$; it is recommended by {\tt emcee} to run the chain at least 50 times $\tau_{int}$, however, as the chain length was increased, $\tau_{int}$ increased as well. This means that the simulation is either divergent or slowly convergent.

After the chain is run, it is flattened; all of the walker’s chains are appended to one single chain. The 16th and 84th quantiles are taken to be the lower and upper confidence intervals respectively. Since MCMC is not an optimizer, these samples are the spread around the resulting median model, rather than the optimized model. The resulting spread of the temperature profile models can be seen in Figures \ref{chandra_model}, \ref{nustar_model}, and \ref{xmm_model}. The spread in the surface brightness model can be seen in Figure \ref{SB_profile}. %\todo{Check after running new MCMC: The confidence interval from the temperature profile contributes $\sim$55\% of the confidence interval in the mass profile, and the gradients of the temperature and density profile together contribute the remaining $\sim$45\%.} 
The median temperature and density model parameters are presented in Tables \ref{pars} and \ref{pars2}, respectively. The parameters can be slightly degenerate (as evidenced by the differing parameters from \chandra,\ \xmm,\ and \nustar\ for the same density profile), so the confidence interval from MCMC is more helpful to view for the models as a whole, rather than for the parameters individually. However, the parameter confidence intervals are also given in Tables \ref{pars} and \ref{pars2}, and a corner plot of the \chandra\ MCMC simulations can be seen in Appendix \ref{cornerplot}.

\begin{table*}
%\centering
\begin{tabular}{ ccccccccc } 
\multicolumn{9}{c}{\large{\textbf{3D Temperature Model Parameters}}}\\
\hline
\hline
 & $T_0$ & $r_t$ & & & & $T_{min}$ & $r_c$ & \\
 & (keV) & (kpc) & $a$ & $b$ & $c/b$ & (keV) & (kpc) & $a_c$ \\
\hline
\chandra:\ & & & & & & & & \\
Fixed $T_{cool}$ & 11.62 & 97.86 & -0.24 & 5.00$^*$ & 0.09 & 4.20 & 129.00 & 1.60 \\
Free $T_{cool}$ & 8.0$^{+0.2}_{-0.2}$ & 260$^{+90}_{-60}$ & -0.11$^{+0.02}_{-0.02}$ & 4.5$^{+0.3*}_{-0.4}$ & 0.05$^{+0.02}_{-0.01}$ & 5.03 & 73.70 & 10.00 \\
\hline
\nustar:\ & & & & & & & & \\
Fixed $T_{cool}$ & 9.29 & 44.38 & -1.00 & 5.00$^*$ & 0.22 & 4.20 & 129.00 & 1.60 \\
Free $T_{cool}$ & 7.3$^{+0.2}_{-0.2}$ & 310$^{+50}_{-40}$ & -0.22$^{+0.01}_{-0.01}$ & 4.4$^{+0.3*}_{-0.7}$ & 0.08$^{+0.03}_{-0.02}$ & 1.32 & 32.55 & 10.00 \\
\hline
\xmm:\ & & & & & & & & \\
Fixed $T_{cool}$ & 9.03 & 46.97 & -0.41 & 5.00$^*$ & 0.11 & 4.20 & 129.00 & 1.60 \\
Free $T_{cool}$ & 7.0$^{+0.2}_{-0.2}$ & 630$^{+200}_{-55}$ & -0.12$^{+0.01}_{-0.01}$ & 4.4$^{+0.3*}_{-0.8}$ & 0.12$^{+0.10}_{-0.05}$ & 3.29 & 28.37 & 10.00 \\
\hline
\end{tabular}
\tablenotetext{*}{5.00 is the maximum value allowed by the fit.}
\caption{The 3D temperature model parameters for \chandra,\ \nustar,\ and \xmm\ both when the $T_{cool}$ parameters are fixed to those given in \citet{vik06_2}, and when they are free to be fit. The fixed $T_{cool}$ parameters are found with a chi-squared fit. The free $T_{cool}$ parameters were first found with a chi-squared fit, then run through MCMC simulations with $T_{min}$, $r_c$, and $a_c$ held constant; excluding these last 3 parameters, the values presented here are the median values from the MCMC simulations. Confidence intervals are thus not given for the last 3 parameters of the free $T_{cool}$ fit or the fixed $T_{cool}$ fit, since they were not simulated via MCMC. Note that it is more informative to view the confidence intervals on the models as a whole, which can be seen in Figures \ref{chandra_model}, \ref{nustar_model}, \ref{xmm_model}, and \ref{SB_profile}.}
\label{pars}
\end{table*}

\begin{table*}
%\centering
\begin{tabular}{ ccccccc } 
\multicolumn{7}{c}{\large{\textbf{Density Model Parameters}}}\\
\hline
\hline
 & $n_1$ & $r_1$ & & $n_2$ & $r_2$ &  \\
 & $(cm^{-3})$ & (kpc) & $\beta_1$ & $(cm^{-3})$ & (kpc) & $\beta_2$ \\
\hline
\chandra:\ & & & & & & \\
Fixed $T_{cool}$ & $1.39\times10^{-4}$ & 59.73 & 0.67 & $3.42\times10^{-5}$ & 166.82 & 0.69 \\
Free $T_{cool}$ & $1.41^{+0.08}_{-0.07}\times10^{-4}$ & 59$^{+4}_{-4}$ & 0.67$^{+0.06}_{-0.04}$ & $3.5^{+0.5}_{-0.4}\times10^{-5}$ & 160$^{+10}_{-10}$ & 0.68$^{+0.02}_{-0.02}$ \\
\hline
\nustar:\ & & & & & & \\
Fixed $T_{cool}$ & $1.46\times10^{-4}$ & 58.56 & 0.66 & $3.38\times10^{-5}$ & 166.74 & 0.68 \\
Free $T_{cool}$ & $1.48^{+0.06}_{-0.06}\times10^{-4}$ & 56$^{+3}_{-3}$ & 0.64$^{+0.03}_{-0.03}$ & $3.23^{+0.5}_{-0.4}\times10^{-5}$ & 170$^{+10}_{-10}$ & 0.69$^{+0.03}_{-0.02}$ \\
\hline
\xmm:\ & & & & & & \\
Fixed $T_{cool}$ & $1.40\times10^{-4}$ & 60.10 & 0.67 & $3.37\times10^{-5}$ & 168.24 & 0.69 \\
Free $T_{cool}$ & $1.41^{+0.08}_{-0.07}\times10^{-4}$ & 59$^{+5}_{-4}$ & 0.67$^{+0.05}_{-0.04}$ & $3.5^{+0.5}_{-0.5}\times10^{-5}$ & 160$^{+10}_{-10}$ & 0.68$^{+0.02}_{-0.02}$ \\
\hline
\end{tabular}
\caption{The density model parameters for \chandra,\ \nustar,\ and \xmm\ both when the $T_{cool}$ parameters are fixed to those given in \citet{vik06_2}, and when they are free to be fit. Though only the Chandra surface brightness profile is fit for the density model, the fit also depends on the temperature model fit, thus the density parameters are slightly different for each profile. The fixed $T_{cool}$ parameters are found with a chi-squared fit. The free $T_{cool}$ parameters were first found with a chi-squared fit, then run through MCMC simulations; thus, confidence intervals are only given for the free $T_{cool}$ fits. Note that it is more informative to view the confidence intervals on the models as a whole, which can be seen in Figures \ref{chandra_model}, \ref{nustar_model}, \ref{xmm_model}, and \ref{SB_profile}.}
\label{pars2}
\end{table*}

\subsection{Calculating Mass From Temperature and Density Profiles}
Relaxed galaxy clusters are assumed to be in hydrostatic equilibrium (HSE), the condition where the gas pressure in the cluster is balanced with the gravitational force.
From the HSE equation, the total mass can be estimated within radius $r$ \citep{vik06_2}:
\begin{equation}
    M_{HSE}(r) = - \frac{k T(r) r}{G \mu m_p} \left(\frac{d\mathrm{log}n(r)}{d\mathrm{log}r} + \frac{d\mathrm{log}T(r)}{d\mathrm{log}r}\right)\, ,
    \label{eq:hse}
\end{equation}
where k, G, $\mu$, and $m_p$ are the Boltzmann constant, gravitational constant, mean molecular weight, and the mass of a proton, respectively. Assuming the ICM is an ionized plasma, the mean molecular weight is 0.5954 \citep{vik06_2}. 
The mass profile depends on radius, X-ray temperature as a function of radius, $T(r)$, and density as a function of radius, $n(r)$. 
%The density in equation \ref{density_equation}, $n_e(r)$, is converted to $n_H(r)$ via re-normalizing by the inverse of the factor that was subsumed
%\todo{how is nH calculated from ne?-- I'm pretty sure it's that normalization thing I have under utilities}
%Together, the value is $-3.71 \times 10^{13} M_{\odot}$.

However, any sources of nonthermal pressure support, such as turbulence and bulk motions due to mergers, can result in an underestimate of hydrostatic mass. \citet{pearce19} state that nonthermal pressure is expected to contribute up to 30\% of the total pressure. They simulate a sample consisting of 45 clusters in the mass range of $8\times10^{13}<M_{500} \left[ M_{\odot} \right] <2\times10^{15}$ in various dynamical states to test the contribution of nonthermal pressure support. They apply a correction to the hydrostatic mass equation (\ref{eq:hse}):

\begin{equation}
    M_{corr}(r)=\frac{1}{1-\alpha}\left[M_{HSE}(r)-\frac{\alpha}{1-\alpha}\frac{kT(r)r}{G\mu m_p} \frac{d\mathrm{log}\alpha}{d\mathrm{log}r}\right]
    \label{eq:corr}
\end{equation}
where $\alpha$ is:
\begin{equation}
    \alpha(r)=1-A\left(1+\mathrm{exp}\left[-\left(\frac{r/r_{500}}{B}\right)^C\right]\right)
\end{equation}
and $A=0.45$, $B=0.84$, $C=1.63$ \citep{nelson2014}, and $r_{500}$ is the radius at which the cluster has an overdensity of 500.

\citet{ettori2021} calculate another model for nonthermal pressure support and apply it to the X-COP galaxy clusters. They find that the constraints on the ratio of nonthermal pressure to total pressure are between 0 and $\sim$20\% at $r_{500}$ for all clusters in the sample except Abell 2319 at $\sim$54\%; the contribution of nonthermal pressure support in Abell 2319 is expected to be higher because it is a merging cluster.

Since the contribution of nonthermal pressure support increases with mass, Eqn. \ref{eq:corr} is an average solution. However, A478 is within the range of masses included in the simulations, so we have applied this correction in this work; the masses calculated with Eqn. \ref{eq:hse} ($M_{HSE}$) as well as those corrected for nonthermal pressure support ($M_{corr}$) are reported in Table \ref{mass}. We find that it results in masses around $13\%$ larger at $r_{2500,Chandra}$ than without nonthermal pressure support.

From the nonthermal pressure support corrected data we find $r_{2500,Chandra}=589\pm14$ and $r_{500,Chandra}$ to be $\sim$1514 kpc, and use these values to calculate the masses from all three telescopes.

\subsection{Mass Profiles}

\begin{figure}[h]
 \centering
 \includegraphics[scale=0.65]{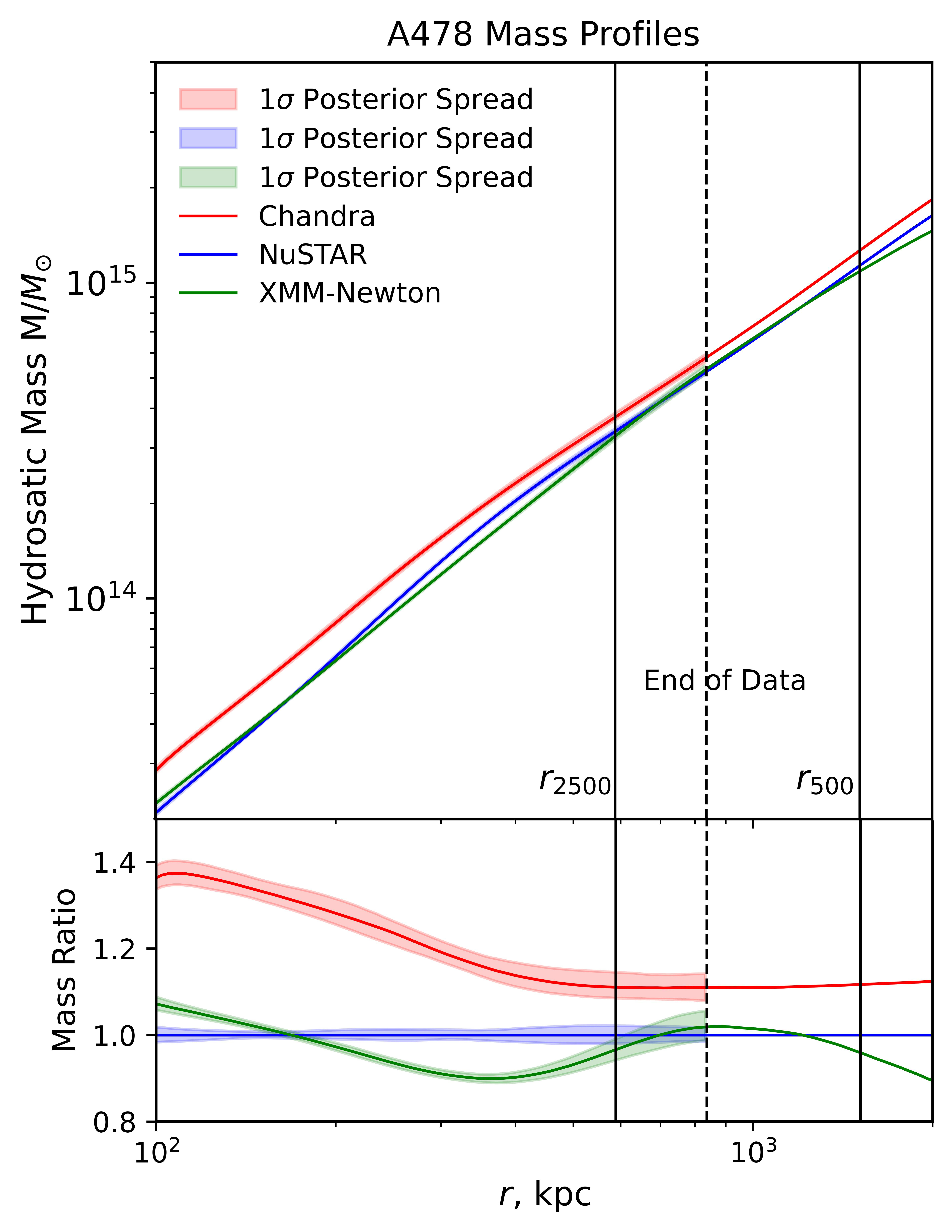}
 \caption{The hydrostatic mass profiles corrected for nonthermal pressure support. The $1\sigma$ posterior spread is propagated from the density and temperature posterior spreads found via MCMC (see Section \ref{sec:MCMC}). The radii $r_{2500}$ and $r_{500}$ are calculated with respect to the nonthermal pressure support corrected \chandra\ data, and are found to be $589 \pm 14$ kpc and $\sim$1514 kpc, respectively. The dashed line shows the end of the largest extracted profile (\xmm) at $\sim$837 kpc. In the top section, the total profiles are shown. In the bottom section, the ratios of the \chandra\ mass profile to the \nustar\ mass profile as well as the \xmm\ mass profile to the \nustar\ mass profile are shown, with the \nustar\ mass profile as a reference.}
 \label{mass_profile}
\end{figure}

The resulting nonthermal pressure support corrected \chandra,\ \xmm,\ and \nustar\ mass profiles are presented in Figure \ref{mass_profile}. As expected, the \chandra\ and \nustar\ masses differ by $\sim$10\% at $r_{2500,Chandra}$, while the difference between \xmm\ and \nustar\ at $r_{2500,Chandra}$ is $\sim$4\%; these differences are the same whether comparing $M_{HSE}$ or $M_{corr}$.

The \chandra\ mass profile is largest due to having the hottest overall temperature. The \chandra\ 3D temperature model has the shallowest slope from 100 kpc to $\sim$350 kpc, which results in the mass increasing more slowly; this is why the \chandra\ mass profile comes closer to the \nustar\ mass profile before $\sim$350 kpc. The \chandra\ and \nustar\ 3D temperature profiles have a similar peak at $\sim$350 kpc, and a similar decreasing slope outside $\sim$350 kpc, resulting in their mass profiles having a similar slope outside $\sim$350 kpc. \xmm\ has the lowest overall temperature, resulting in the smallest mass profile. However, \xmm’s\ 3D temperature profile has a hotter core than \nustar's,\ so the \xmm\ mass profile starts at a higher mass than \nustar\ at 100 kpc. From 100 kpc to $\sim$350 kpc, \xmm’s\ 3D temperature profile has a shallower slope than \nustar’s,\ resulting in \xmm’s\ mass profile increasing more slowly; this causes the \xmm\ mass profile to cross \nustar’s\ and become smaller. Then, from $\sim$350 kpc to $\sim$600 kpc \xmm’s\ 3D temperature profile continues increasing while \nustar’s\ reaches its peak and begins decreasing. This causes the \xmm\ mass profile to increase more quickly than the \nustar\ mass profile, once again crossing it, and end up just above the \nustar\ mass profile at $\sim$837 kpc.

The confidence intervals shown in Fig. \ref{mass_profile} are the result of calculating the mass for each set of parameters simulated by the MCMC simulation, then calculating the 16th and 84th quantiles. The \chandra,\ \nustar,\ and \xmm\ masses at $r_{2500,Chandra}$ and $r_{500,Chandra}$ are given in Table \ref{mass}. Due to our largest profile ending around $\sim$837 kpc, $r_{500,Chandra}$, as well as all masses calculated at this radius, are an extrapolation of our data and thus are not presented with confidence intervals. We compare to masses found in other works as well. Since each work measures the overdensities at different radii, these radii are also given in Table \ref{mass}.

\begin{table*}
%\centering
\begin{tabular}{ cccccc }
\multicolumn{6}{c}{\large{\textbf{Mass of A478}}}\\
\hline
\hline
 & & $M_{2500}$ & $r_{2500}$ & $M_{500}$ & $r_{500}$ \\
 & & ($10^{14}$ $M_\odot$) & (kpc) & ($10^{14}$ $M_\odot$) & (kpc) \\
\hline
{\multirow{3}{*}{\rotatebox[origin=c]{90}{$M_{HSE}$}}} & \chandra\ & $3.28^{+0.10}_{-0.07}$ & $589\pm14^{*}$ & 8.00$^{\dagger}$ & $1514^{*}$ \\ 
& \nustar\ & $2.96^{+0.06}_{-0.06}$ & $589\pm14^{*}$ & 7.17$^{\dagger}$ & $1514^{*}$ \\ 
%& 18.45
& \xmm\ & $2.86^{+0.07}_{-0.07}$ & $589\pm14^{*}$ & 6.87$^{\dagger}$ & $1514^{*}$ \\ 
\hline
{\multirow{3}{*}{\rotatebox[origin=c]{90}{$M_{corr}$}}} & \chandra\ & $3.77^{+0.12}_{-0.09}$ & $589\pm14^{*}$ & 12.79$^{\dagger}$ & $1514^{*}$ \\ 
& \nustar\ & $3.39^{+0.07}_{-0.07}$ & $589\pm14^{*}$ & 11.45$^{\dagger}$ & $1514^{*}$ \\ 
%& 18.45
& \xmm\ & $3.28^{+0.08}_{-0.08}$ & $589\pm14^{*}$ & 10.98$^{\dagger}$ & $1514^{*}$ \\ 
\hline
 & \chandra\ \citep{vik06_2} & $4.12\pm0.26$ & $\sim0.4r_{500}$ & $7.68\pm1.01$ & $1337\pm58$ \\ 
\hline
 & \xmm\ \citep{arnaud2005} & $3.12\pm0.31$ & $\sim0.3r_{200}^\ddagger$ & $7.57\pm1.11$ & $1348\pm64$ \\  
\hline
 & \chandra\ \citep{mahdavi07} & $4.9\pm0.9$ & $680\pm70$ & $9.9\pm2.6$ & $1500\pm200$\\ 
% & $13\pm4$ 
 & \xmm\ \citep{mahdavi07} & $3.3\pm0.2$ & $590\pm20$ & $8.2\pm1.0$ & $1400\pm100$ \\ 
\hline
 & \chandra\ \citep{Mantz_2016} & $3.69\pm0.15$ & $620\pm8$ & - & -\\  
\hline
 & \chandra\ \citep{wulandari19} & $2.0\pm0.5$ & $570\pm80^{\S}$ & - & -\\  
\hline
 & Sunyaev Z'eldovich Effect \citep{comis2011} & $2.8\pm1.1$ & $570\pm80$ & - & -\\  
\hline
\end{tabular}
\tablenotetext{*}{The masses of \nustar\ and \xmm\ are calculated at $r_{2500,Chandra}$ and $r_{500,Chandra}$ as well, to make comparison as simple as possible.}
\tablenotetext{\dagger}{The extracted temperature and surface brightness profiles only reach $\sim$250 kpc past $r_{2500,Chandra}$, thus the masses at $r_{500,Chandra}$ are an extrapolation.}
\tablenotetext{\ddagger}{\citet{arnaud2005} report $r_{200}=2060\pm618$ kpc.}
\tablenotetext{\S}{\citet{wulandari19} adopt the value of $r_{2500}$ found by \citet{comis2011}.}
\caption{The \chandra,\ \nustar,\ and \xmm\ total masses found in this work at $r_{2500,Chandra}=589\pm14$ and $r_{500,Chandra}\sim1514$ kpc both calculated from the hydrostatic mass equation directly ($M_{HSE}$) and corrected for nonthermal pressure support ($M_{corr}$). Masses from other works are shown for reference; since each work calculates a different value for $r_{2500}$ and $r_{500}$, these radii are given in the third and fifth columns, respectively. Additionally, the cosmology assumed in these works varies, resulting in varying values for the critical density, $\rho_{c}$(z), and thus differences in the relation between $r_{\Delta}$ and $M_{\Delta}$. All uncertainties shown are $1\sigma$.}

%total masses found in this work at $r_{2500}=570\pm80$ kpc \citep{comis2011}, $r_{500}=1337\pm58$ kpc \citep{vik06_2}, and $r_{200}=2060\pm112$ kpc \citep{arnaud2005}. Also included are \chandra,\ \nustar,\ \xmm,\ and Sunyaev Z'eldovich Effect masses from other works. \citet{Mantz_2016} find $r_{2500}=620\pm8$ kpc. \citet{mahdavi07} find $r_{2500}=0.68\pm0.07$ Mpc, $r_{500}=1.5\pm0.2$ Mpc, and $r_{200}=2.2\pm0.4$ Mpc for \chandra\ and $r_{2500}=0.59\pm0.02$ Mpc, $r_{500}=1.4\pm0.1$ Mpc, and $r_{200}=2.1\pm0.2$ Mpc for \xmm.\ All other radii are either not specified or are the same as used in this work. All masses are presented in units of $10^{14}$ $M_\odot$.}
%\todo{wulandari also uses the comis r2500, Mantz uses r2500=620\pm8, Mahdavi (Chandra) uses r2500=0.68\pm0.07 Mpc, r500=1.5\pm0.2 Mpc, and 2.2\pm0.4 Mpc,  Mahdavi (XMM) uses r2500=0.59\pm0.02 Mpc, r500=1.4\pm0.1 Mpc, and 2.1\pm0.2 Mpc,}}
\label{mass}
\end{table*}

\section{Discussion}\label{sec:discussion}

%\AT {stating the problem ie"Temperature differences due to cross calibration issues affect the mass measurements of galaxy clusters." and then describe your cluster literature and your results} 
%\todo{need to include ideas for why my chandra and xmm masses are different than chandra and xmm masses from other papers. Also, should I include a weak-lensing mass?}

%The differences in temperature measured by \chandra,\ \nustar,\ and \xmm\ due to cross-calibration \todo{I think you first have to state that the observed differences are actually due to cross calibration. What other factors could influence here? Multiphase gas (probably not because Nustar would be higher). PSF effects?} issues affect the galaxy cluster mass measurements. 
We find that the temperature profile of A478 measured by \nustar\ is $\sim$11\% cooler than the \chandra\ temperature profile on average. This results in the \nustar\ mass at $r_{2500,Chandra}$ being $\sim$10\% lower than the \chandra\ mass, and $\sim$4\% higher than the \xmm\ mass. \citet{wallbank22} find that, for a sample of 8 galaxy clusters, \nustar\ temperatures are $\sim$10\% and $\sim$15\% lower than \chandra\ temperatures in the broad and hard bands, respectively. They discuss the effect the uncertainty in the background modeling might have on the temperatures and find that the signal-to-noise was high enough in their sample that the temperature fits were insensitive to the background modeling. They also find that, since the \chandra\ temperatures remain systematically higher than \nustar\ when limited to the hard band, this discrepancy is not due to any factors that affect soft band modeling, such as absorption along the line of sight and ACIS contamination.

The difference in the shape of the temperature profile of A478 between \chandra\ and \xmm\ is explained by \citet{vik06_2} to be an incomplete correction of \xmm's\ PSF. Though \nustar\ does have a larger PSF than both \chandra\ and \xmm,\ this was corrected using the shape of the PSF to determine the scatter of emission from one annulus to another and creating ARFs to account for it. While we do not quantify any systematic effects of {\tt nucrossarf} here, it does not appear to have any bias, as forthcoming work will show. Even after this correction, the \nustar\ profile is in better agreement with \xmm\ than \chandra.\ Though \xmm's\ PSF is not corrected here, the annular regions are large enough that we believe the effect of it to be negligible.

%\AT {More on calibration from Dan's proposal} 
%From NuSTAR Proposal: "Temperature measurements can be bandpass dependent, especially if the spectrum is not isothermal (Mazzotta et al. 2004), and slight soft band calibration issues can conspire with the Galactic column density nH to bias temperature estimates. These effects are worse for massive or hot clusters because the exponential turnover in the bremsstrahlung continuum falls outside the bandpass of Chandra or XMM-Newton, limiting the constraining power of their spectra."

Bandpass can affect temperature measurements; at softer energies, temperature is determined mainly by the power law-like slope of the bremsstrahlung spectrum, which can be biased by other sources of emission or mischaracterized background or absorption due to Galactic column density. %\todo{
In this work, the broad bands of each instrument were used to determine temperature; \chandra\ spectra were fit from 0.8--9 keV, \xmm\ spectra from 0.4--11 keV, and \nustar\ spectra from 3--15 keV.
%} 
Thus, \nustar\ is less susceptible to these issues than \chandra\ and \xmm, being more sensitive to the exponential turnover of the bremsstrahlung spectrum, where temperature can be more accurately measured.

To extract the projected temperature profile, the ICM is assumed to be isothermal in each annulus, but this is not the case (see Fig. \ref{tmap}). Given that \nustar\ is more sensitive to higher energies than \chandra\ and \xmm, and thus would be more sensitive to the higher temperature components of the gas, \nustar\ would be expected to measure higher temperatures than both \chandra\ and \xmm.\ However, we find that the \nustar\ temperature profile for A478 is lower than \chandra\ and not much higher than \xmm.\
%\AT{you can site my work (tumer2022 in the name.bib, https://iopscience.iop.org/article/10.3847/1538-4357/ac61de/pdf Table 4) here where I found nustar lower than XMM}

\citet{sanderson05} simulated multiphase gas with four temperature components (6, 6.5, 7.5, and 8 keV) using both the \chandra\ and \xmm\ spectral responses and background spectra of A478. The simulations had a Galactic absorption of $2.7\times10^{21}$ cm$^{-2}$ and an abundance of 0.3 Z$_{\odot}$. They used a single absorbed {\tt MEKAL} model to fit the spectra with both the narrow (6.0--6.8 keV) and broad (0.7--7.0 keV) bands of both \chandra\ and \xmm.\ They find that the fits do not show any significant disagreement between \chandra\ and \xmm.\ With another set of simulations with two temperatures (5 and 12 keV) and different abundances, they find that when the hotter phase has a higher metallicity, the narrow band temperature increases for both \chandra\ and \xmm,\ and a lower metallicity for the hotter phase decreases the narrow band temperature for both telescopes. Again, there is no significant disagreement between the \chandra\ and \xmm\ temperatures. Thus, they conclude that the discrepancy between \chandra\ and \xmm\ cannot be entirely attributed to nonisothermality in the ICM. 

Similarly, \citet{schellenberger2014} simulated spectra for \chandra\ ACIS-I and all three \xmm\ instruments. The cold temperature components were 0.5, 1, and 2 keV, and the hot temperature component was varied from 3 to 10 keV. The hot component had a fixed abundance of 0.3 $Z_\odot$, while the cold components had 0.3, 0.5, and 1 $Z_\odot$, respectively. The spectra were fit with single temperature models, with the abundance free to vary. They find that when the cold temperature component is $\sim$2 keV, the fit is good, regardless of the amount of cold gas. The temperatures only begin to differ significantly when the cold component is $\sim$0.5 keV. Thus, they conclude that multiphase gas is not enough to explain the differences between the \chandra\ and \xmm\ temperatures.

%\todo{Gerrit: We did something very similar in Schellenberger+15. Similar conclusion, you would need a very strong cold component, and it is impossible to explain all the differences (ACIS/PN + ACIS/MOS + PN/MOS) at the same time.}

\nustar\ is also less sensitive to absorption than Chandra and \xmm\ due to its lack of sensitivity below 3 keV \citep{rojas2021}, meaning the complicated absorption along the line of sight of A478 has less of an effect on temperature measurements. \citet{tumer2023} find that leaving the $N_H$ parameter free for a global \nustar\ fit of CL 0217+70 results in unphysical behavior, demonstrating \nustar's\ insensitivity to absorption.
%\AT {More on absorption from my paper}

The \xmm\ temperature profile used by \citet{arnaud2005} to find the mass of A478 is given by \citet{pointe04}. Their profile peaks around 3.66--4.54\arcmin\ or $\sim$400 kpc and is 6.91 keV, which is around 0.5 keV hotter than found in this work. \citet{deplaa04} find an \xmm\ profile that peaks around 2.0--3.0\arcmin\ or $\sim$250 kpc at 6.76 keV, around 0.3 keV higher than ours. \citet{sanderson05} find a profile that also peaks around $\sim$250 kpc (123--175\arcsec), and their peak temperature is 7.06 keV, which is $\sim$0.6 keV higher than ours. \citet{bourdin07} measure their peak temperature of 7.5 keV around 3.5\arcmin or $\sim$340 kpc; this is closest to our peak radially, but a little over 1 keV hotter.

Our \xmm\ temperature profile of A478 agrees fairly well with other recent works; though it is cooler than each of the profiles discussed here, the peak of the profiles line up radially. The cooler temperatures are likely due to the recent \xmm\ calibration update, which brought it into better agreement with \nustar.\ This update included the addition of the {\tt applyabsfluxcorr} keyword, which we make use of in this work. The effect this correction has on temperature can be seen in Appendix \ref{xmm_fluxcorr}. Our \xmm\ temperature profile compared to other works can be seen in Appendix \ref{other_temps}.

Our \chandra\ temperature profile is cooler than found by \citet{Vik05} by $\sim$1 keV, though their peak lines up with ours at around 250--300 kpc. \citet{sanderson05} also find a temperature profile that seems to peak around 175--250$\arcsec$ or $\sim$300 kpc, at 8.12 keV, which is hotter than our peak by $\sim$0.3 keV. The peak temperature found by \citet{Mantz_2016} is around 8 keV, thus $\sim$0.2 keV hotter than ours, and also appears to peak around 300 kpc. Thus our \chandra\ profile, while cooler than those found in other works, peaks at around the same radius and has the same general shape. Our \chandra\ profile compared to that of \citet{sanderson05} can be seen in Appendix \ref{other_temps}.

Differences in analysis such as how the data is reduced, which models are used to fit the spectra, and calibration updates will cause differences in measured temperature. The \chandra\ and \xmm\ temperature profiles from this work compared to other works whose temperature profiles were given in tables can be seen in Appendix \ref{other_temps}.

Our final masses at $r_{2500,Chandra}$ and $r_{500,Chandra}$ are given in Table \ref{mass}. We give the 68\% confidence interval for the mass at $r_{2500}$, but not for the extrapolated masses at $r_{500}$ since our data does not extend that far. We've included masses and the radii at which they're measured from other works as well.

At $r_{2500,Chandra}$ our \chandra\ mass is $\sim$8\% smaller than found by \citet{vik06_2} at their measured $r_{2500}$, which is expected since our temperature profile is cooler. Our \chandra\ mass at $r_{2500,Chandra}$ is also $\sim$23\% smaller than found by \citet{mahdavi07}, at their $r_{2500}$ who jointly fit X-ray, Sunyaev-Zel'dovich effect (SZE), and weak lensing data to calculate the mass. The \chandra\ mass at their $r_{2500}$ found by \citet{Mantz_2016} is 2$\%$ smaller than ours at $r_{2500,Chandra}$, and the mass found by \citet{wulandari19}, who adopt the value of $r_{2500}$ found by \citet{comis2011}, is $\sim$47\% smaller than ours. Our \xmm\ mass at $r_{2500,Chandra}$ is $\sim$5\% larger than found by \citet{arnaud2005} at their $r_{2500}$, and $\sim$1\% larger than found by \citet{mahdavi07} at their $r_{2500}$. Our \chandra,\ \xmm,\ and \nustar\ masses at $r_{2500,Chandra}$ are $\sim$34\%, $\sim$17\%, and $\sim$21\% larger than the x-ray calibrated SZE mass at their $r_{2500}$ found by \citet{comis2011} respectively. %\todo{\citet{mahdavi07} get a difference of over 30\% between the \chandra\ and \xmm\ masses at $r_{2500}$.}
%, which decreases to 8\% at $r_{200}$. The difference between our \chandra\ and \xmm\ masses also decreases from $\sim$9\% at $r_{2500,Chandra}$ to $\sim$8\% at $r_{500,Chandra}$.

Our \nustar\ mass at $r_{2500,Chandra}$ is smaller than all but one of \chandra\ masses presented (at their respective $r_{2500}$) by up to $\sim$31\%; it is $\sim$41\% larger than the \chandra\ mass found by \citet{wulandari19} at their $r_{2500}$. Our \nustar\ mass at $r_{2500,Chandra}$ is $\sim$3$\%$ larger than the \xmm\ mass found by \citet{mahdavi07} at their $r_{2500}$, and $\sim$9\% larger than the \xmm\ mass found by \citet{arnaud2005} at their $r_{2500}$.

Assuming \nustar\ is more accurate, cluster masses previously measured with \chandra\ could be $\sim$10\% too large. The effect will be strongest on $\sigma_8$, which is sensitive to the high cluster mass end of the mass function. With the HIFLUGCS sample of clusters, \citet{schellenberger2017} find that the smaller \xmm\ masses result in a lower $\sigma_8$ than \chandra\ masses. The smaller \nustar\ masses found in this work would result in lower $\sigma_8$ values than the \chandra\ masses as well. These results, however, are based entirely on one cluster; Abell 478. This analysis will be repeated for three other clusters also recently observed with \nustar\ to determine whether this difference in mass is systematic or unique to this cluster.

%\todo{Gerrit: OmegaM is sensitive to the overall cluster number density as a fucntion of mass, the effect should be weaker here, and more strong on sigma8, which is sensitive to the exponential cut off at high cluster mass. You could cite my other 2015 paper \url{http://dx.doi.org/10.1093/mnras/stx1583} "the difference between Chandra and XMM–Newton masses is increasing with mass, which leads to lower sigma8 of the XMM–Newton masses with respect to Chandra. We confirm the previous results that the overall shift in the omegam–sigma8 plane cannot explain the difference between cosmological constraints of Planck primary CMB anisotropies and SZ, and also that for the present study, the shift is smaller than the statistical uncertainty."}

%\AT {Compare temp with Bourdin 2007. Compare mass with Wulandari 2019. Use the following from bourdin 2009 for nh affect " it is located in a particular sky region charac- terised by strong angular variations of the neutral hydrogen column density, Nh, across the cluster field of view. If not treated properly, these strong Nh variations may introduce spurious thermal features that cannot be easily disentan- gled from the real ones."}

\begin{acknowledgments}

This research was supported by the NASA grant 80NSSC21K0075. This research has made use of the \nustar\ Data Analysis Software (NuSTARDAS)\footnote{\url{https://heasarc.gsfc.nasa.gov/docs/nustar/analysis/}} jointly developed by the ASI Space Science Data Center (SSDC, Italy) and the California Institute of Technology (Caltech, USA). The data analysis software HEASoft\footnote{\url{https://heasarc.gsfc.nasa.gov/lheasoft/}}, maintained by NASA’s HEASARC, was used in the analysis of both the \chandra\ and \nustar\ data, and the \chandra\ X-ray Center’s software package CIAO\footnote{\url{https://cxc.cfa.harvard.edu/ciao/}} was used in analyzing the \chandra\ data. {\textit XMM}--ESAS\footnote{\url{https://heasarc.gsfc.nasa.gov/docs/xmm/xmmhp\_xmmesas.html}}, developed at Goddard Space Flight Center with cooperation with the \xmm\ Science Operations Center, was used in the analysis of the \xmm\ data. The data was obtained from the \chandra\ Data Archive, the \xmm\ Science Archive, and the \nustar\ mission, which is led by the California Institute of Technology (Caltech, USA), managed by the Jet Propulsion Labratory, and funded by NASA.

The MCMC tutorial ``MCMC: A (very) Beginnner’s Guide,"\footnote{\url{https://prappleizer.github.io/Tutorials/MCMC/MCMC_Tutorial.html}} written by Imad Pasha was extremely helpful for writing the MCMC code for confidence intervals.

\end{acknowledgments}

\nocite{*}

\bibliography{name}
\bibliographystyle{aasjournal}

\appendix
\section{\chandra\ Full and Hard Bands}

The \chandra\ temperatures for each region were fit in the full band, from 0.8--9.0 keV. \chandra\ is mostly consistent between soft and hard bands. However, since the soft band is sensitive to absorption and the hard band is not, this can be one source of discrepancy between them. The absorption along the line of sight of Abell 478 is complicated and left free in the full band annular \chandra\ temperature fits.

\begin{figure}[h]
 \centering
 \includegraphics[scale=0.35]{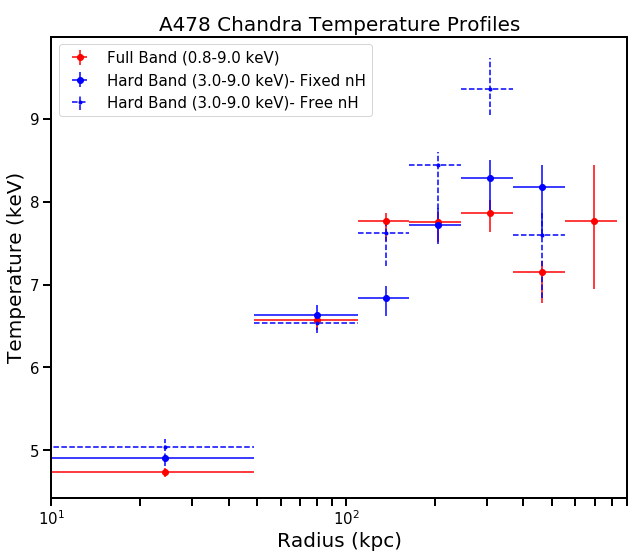}
 \caption{The full (red) and hard (blue) band \chandra\ temperature fits. The solid blue points are the hard band fits with the absorption ($N_H$) fixed to the best-fit value found in the full band fits. The dashed blue points are the hard band fits when the absorption is left free to fit. The last annulus is not included for the hard band fits because the signal-to-noise was low, which prevented the temperature from being constrained.}
 \label{fullvhard}
\end{figure}

Two hard band fits from 3.0 to 9.0 keV are shown in Fig. \ref{fullvhard} in comparison to the full band. When the absorption is fixed to the full band best-fit values, the largest difference is in the 3rd region, where the hard band fit temperature is $\sim$12$\%$ lower than the full band fit. Freeing the absorption in the hard band fits reduces the discrepancy in the 3rd region but also increases the discrepancy in the 5th region to $\sim$19$\%$.

\section{\chandra\ ACIS-S vs. ACIS-I}

\begin{figure}[h]
 \centering
 \includegraphics[scale=0.35]{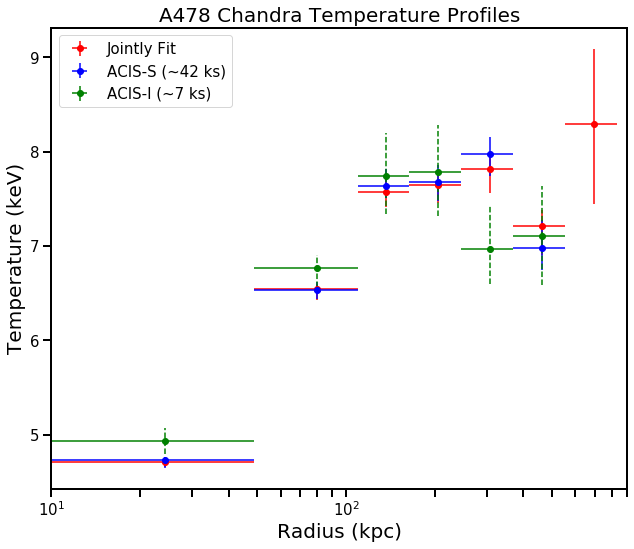}
 \caption{The joint (red), ACIS-S (blue), and ACIS-I (green, dashed) \chandra\ temperature fits. The outermost annulus is not included for the individual ACIS-S and ACIS-I fits because the signal-to-noise was too low to constrain the temperature.}
 \label{acis}
\end{figure}

The two \chandra\ observations used in this work are a $\sim$42 ks ACIS-S observation (1669) and a $\sim$7 ACIS-I observation (6102). The observations were jointly fit for the \chandra\ temperature profile used to calculate the mass. Individual fits of the observations compared to the joint fit are shown in Fig. \ref{acis}. For the individual fits, the absorption was fixed to the best-fit values from the joint fit. The largest difference is in the 5th region, where the ACIS-I fit is $\sim$13$\%$ lower than the joint fit. The 5th region is also the one with the largest discrepancy in the free absorption hard band fit (see Fig. \ref{fullvhard}).

\section{ {\tt applyabsfluxcorr} Correction }
\label{xmm_fluxcorr}

\begin{figure}[h]
 \centering
 \includegraphics[scale=0.35]{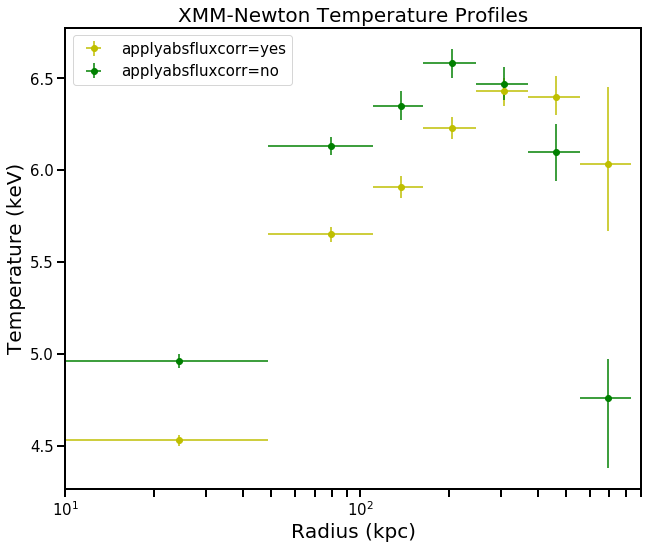}
 \caption{The \xmm\ temperature profiles with {\tt applyabsfluxcorr=yes} (yellow) and {\tt applyabsfluxcorr=no} (green).}
 \label{xmm_fluxcorr_plot}
\end{figure}

The {\tt applyabsfluxcorr} keyword was set to yes for the \xmm\ data reduction in this work. This correction does not correct the flux but increases the $E>4$ keV EPIC effective area. This correction brings \xmm\ and \nustar\ into better agreement. The difference the correction has on the temperature profile can be seen in Fig. \ref{xmm_fluxcorr_plot}; with the correction, the temperature profile is cooler in the first four regions by 8\% on average. It is only cooler by $\sim$1\% in the 5th region, and hotter by $\sim$5\% and $\sim$21\% in the 6th and 7th regions respectively. 

\section{Absorption}
\label{abs}

The absorption parameter ($N_H$) was left free in each annular fit of both \chandra\ and \xmm\ due to the variation of column density with radius. Attempts to fix either the \chandra\ $N_H$ to the \xmm\ best-fit values, or the \xmm\ $N_H$ to the \chandra\ best-fit values resulted in poorer fits overall. The \chandra\ fits with \xmm\ best-fit $N_H$ are hotter by 11\% on average, with the largest difference in the 3rd to last region at 17\% hotter. The last region is not included for the fixed $N_H$ fits because it was unable to constrain the temperature.

\begin{figure}[h]
 \centering
 \begin{minipage}{.45\textwidth}
  \centering
  \includegraphics[scale=0.325]{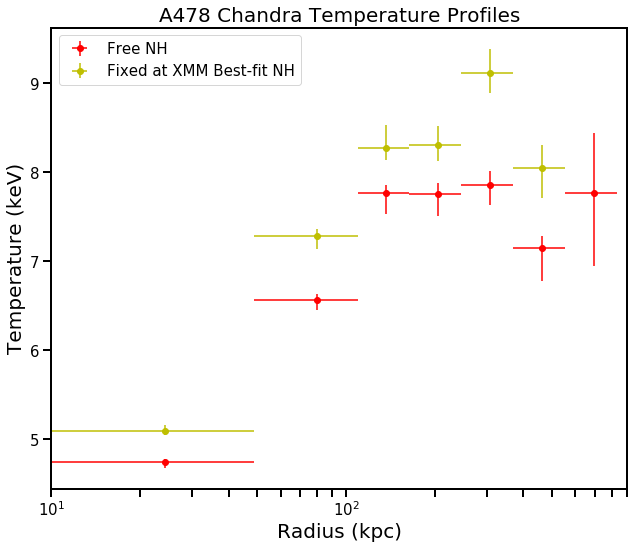}
  \label{chandra_xmmnh}
 \end{minipage}%
 \begin{minipage}{.2\textwidth}
 \end{minipage}%
 \begin{minipage}{.45\textwidth}
  \centering
  \includegraphics[scale=0.25]{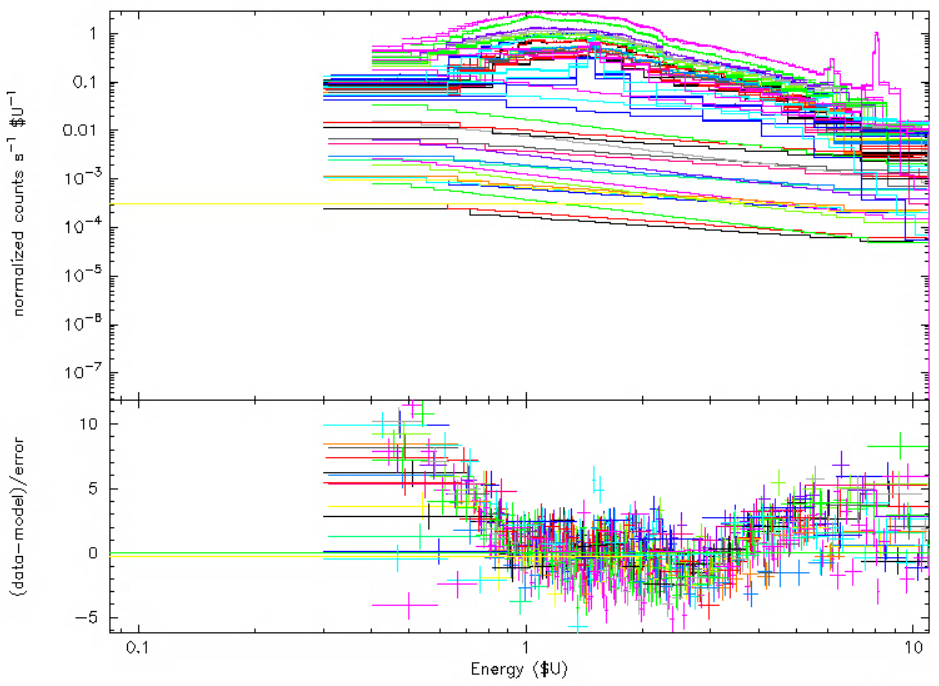}
  \label{xmm_ns}
 \end{minipage}
 \caption{(Left) The \chandra\ temperatures used in this work (red) as well as the \chandra\ temperatures with $N_H$ fixed at the \xmm\ best-fit values (yellow). (Right) The \xmm\ spectra fitted with $N_H$ fixed to the \chandra\ best-fit values.}
 \label{absorption}
\end{figure}

The \xmm\ spectra with $N_H$ fixed to the \chandra\ best-fit values can be seen in Figure \ref{absorption}. It is very poor fit, and thus we do not provide the resulting temperature profile.

\pagebreak

\section{Comparing to other temperature profiles}
\label{other_temps}

\begin{figure}[h] 
 \centering
 \begin{minipage}{.45\textwidth}
  \centering
  \includegraphics[scale=0.35]{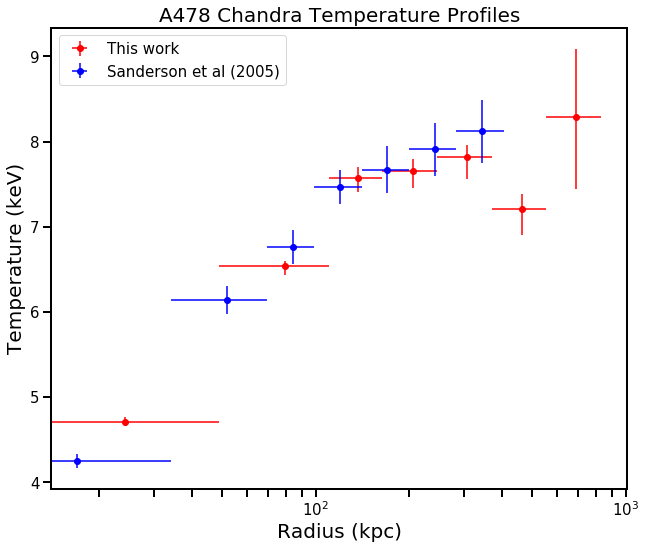}
  \label{other_chandra}
 \end{minipage}%
 \begin{minipage}{.2\textwidth}
 \end{minipage}%
 \begin{minipage}{.45\textwidth}
  \centering
  \includegraphics[scale=0.35]{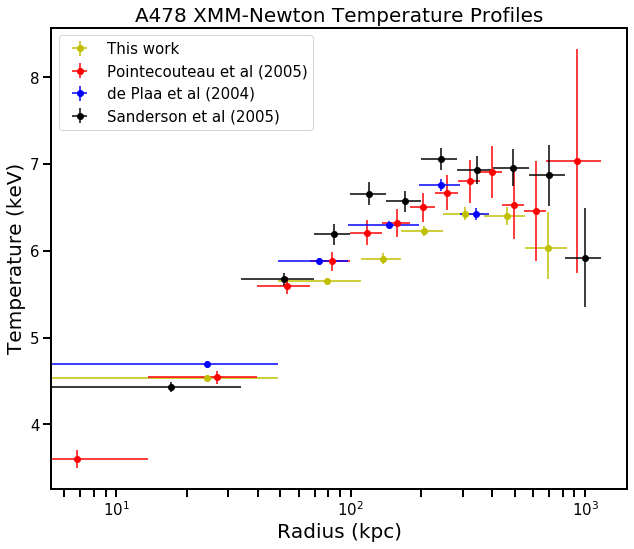}
  \label{other_xmm}
 \end{minipage}
 \caption{(Left) The \chandra\ temperature profiles from this work (red) and found by \citet{sanderson05} (blue). (Right) The \xmm\ temperature profiles from this work (yellow), \citet{pointe04} (red), \citet{deplaa04} (blue), and \citet{sanderson05} (black). Our \xmm\ profile is cooler than found in other works; this is likely due to the application of the {\tt applyabsfluxcorr} correction (see Appendix \ref{xmm_fluxcorr}).}
\end{figure}

\pagebreak

\section{Markov Chain Monte Carlo Corner Plot}
\label{cornerplot}

\begin{figure}[h]
 \centering
 \includegraphics[scale=0.11]{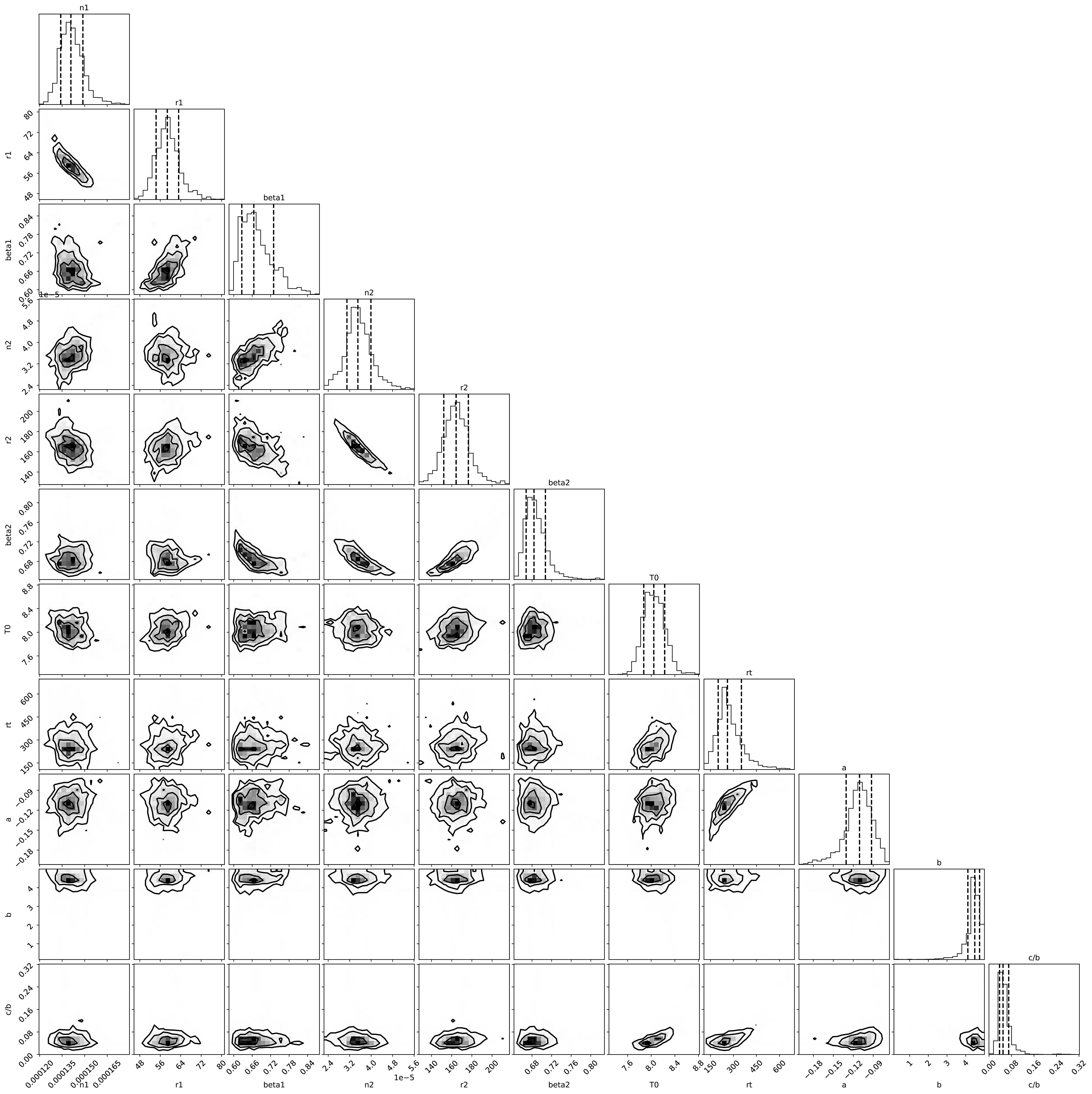}
 \caption{The corner plot of the \chandra\ MCMC simulations (see Section \ref{sec:MCMC}). The parameters are the double-$\beta$ density parameters and the temperature parameters excluding $T_{cool}$. The simulation was only run for 1000 iterations with 50 walkers, making the contours fairly rough. The \nustar\ and \xmm\ corner plots are similar.}
 \label{chandra_cornerplot}
\end{figure}

\end{document}